\documentclass[12pt,preprint]{aastex}

\shorttitle{Sequential star formation in BRC37}
\shortauthors{Ikeda et al.}

\begin{document}


\title{Sequential star formation in a cometary globule (BRC37) of IC1396}

\author{Hisashi Ikeda,\altaffilmark{1}  Koji Sugitani,\altaffilmark{1}  Makoto Watanabe,\altaffilmark{2}  
Naoya Fukuda,\altaffilmark{3} Motohide Tamura,\altaffilmark{4} Yasushi Nakajima,\altaffilmark{4}
Andrew J. Pickles,\altaffilmark{5} Chie Nagashima,\altaffilmark{6} Takahiro Nagayama,\altaffilmark{7} 
Hidehiko Nakaya,\altaffilmark{2} Makoto Nakano,\altaffilmark{8} and Tetsuya Nagata\altaffilmark{7}}

\altaffiltext{1}{Graduate school of Natural Sciences, Nagoya City University, Mizuho-ku,
 Nagoya 467-8501, Japan}
\altaffiltext{2}{Subaru Telescope, National Astronomical Observatory of Japan, Hilo, HI 96720}
\altaffiltext{3}{Department of Computer Simulation, Okayama University of Sience, 1-1 Ridai-cho,
Okayama 700-0005, Japan}
\altaffiltext{4}{National Astronomical Observatory, 2-21-1 Osawa, Mitaka, Tokyo 181-8588, Japan}
\altaffiltext{5}{Las Cumbres Observatory, Goleta CA 93117}
\altaffiltext{6}{Department of Astrophysics, Nagoya University, Chikusa-ku, Nagoya 464-8602, Japan}
\altaffiltext{7}{Department of Astronomy, Kyoto University, Sakyo-ku, Kyoto 606-8502, Japan}
\altaffiltext{8}{Faculty of Education and Welfare Science, Oita University, Oita 870-1192, Japan}

\begin{abstract}
We  have carried out near-IR/optical observations to examine star formation 
toward a bright-rimmed cometary globule (BRC37) facing the exciting star(s) of an HII region (IC1396) 
containing an IRAS source, which is considered to be an intermediate-mass protostar. 
With slitless spectroscopy we detected ten H$\alpha$ emission stars around the globule, six of 
which are near the tip of the globule and are aligned along the direction to the exciting stars.
There is evidence that this alignment was originally toward an O9.5 star, but has evolved to align toward 
a younger O6 star when that formed.
Near-IR and optical photometry suggests that four of these six stars
are low-mass young stellar objects (YSOs) with masses of $\sim0.4$ M$_\sun$. 
Their estimated ages of $\sim 1$ Myr indicate that they were formed at the tip in advance of the formation 
of the IRAS source. 
Therefore, it is likely that sequential star formation has been taking place along the direction 
from the exciting stars toward the IRAS source, due to the UV impact of the exciting star(s). 
Interestingly, one faint, H$\alpha$ emission star, which is the closest to the exciting star(s), 
seems to be a young brown dwarf that was formed by the UV impact in advance of the formation of 
other YSOs at the tip.

\end{abstract}


\section{Introduction}

Small molecular clouds or globules are often seen as small dark 
patches in the peripheries of HII regions, having bright rims 
on the side facing the exciting stars of the HII regions.
These bright-rimmed clouds (BRCs) /globules are 
considered to be pre-existing denser parts in the original molecular 
cloud that formed the HII region, or parts that became denser 
due to the expansion of the HII region. After evaporation of 
their thinner ambient molecular gas, they will be exposed 
directly to the UV radiation from the exciting stars, forming ionization 
fronts (bright rims) on the exciting star sides. 
Because of the high pressure of the bright rim and the geometrical focusing, 
molecular gas could be squeezed and consequently has a higher density to form stars
\citep[e.g.,][]{sand82,bertoldi89,leflo94,williams01,kessel03,miao06,motoyama07}.
Therefore, BRCs are potential sites 
of induced star formation due to compression by ionization/shock fronts.

In fact, many signposts of star formation, i.e., IRAS
point sources, molecular outflows, HH objects, reflection nebulae, 
and so on, have been reported in BRCs \citep[e.g., Table 1 of][]{elme98}.
Our systematic near-IR imaging of the IRAS-associated 
BRCs of \cite{sugitani91} revealed that many of them are associated 
with small young clusters or aggregations, some of which 
have asymmetric distributions biased toward the exciting-star side along the axes of BRCs. 
This indicates that these young stars formed prior to 
the formation of the IRAS sources (protostar-like objects), i.e., propagation 
of star formation along the axis of BRC as the ionization/shock 
front advances outward of the HII region \citep{sugitani95}.
Recently, this sequential star formation was more clearly confirmed 
in BRC14 \citep[SFO14;][]{matu06}  and BRC38 \citep[SFO38;][]{getm07},  
and the precursor of such star formation was reported in BRC75 \citep[SFO75;][]{urq07}.

Three-dimensional  (3D) simulations imply that some radiationally imploded globule repeat leaving
small cores decoupled from itself, which collapse after passage of the ionization/shock front  
and form stars, i.e., sequential star formation in the globule \citep{kessel03}.

BRC37 (SFO37) is a small cometary globule with a bright rim in the southern 
periphery of IC1396 \citep[$d\sim$750 pc;][]{matthews79}.  
It is located $\sim$12 pc from the main exciting star (HD206267, O6) 
and the size of this globule is $\sim$0$'$.6 wide and $\sim$3$'$ long in the optical (Fig. \ref{fig1}).  
IRAS 21388+5622 ($L\sim155 L_\sun$) is located at the tip of the globule. 
It is considered to be an A star \citep{sch91} having a molecular outflow 
with a dynamical age of $\sim$0.3 Myr \citep{duvert90}.  
Recently, \cite{morg08} reported that the luminosity of the internal source of this globule was 
estimated to be $63 L_\sun$, corresponding to a spectral type of B9.

Our interferometric $^{13}$CO observations clearly showed 
two tails of blueshifted velocities that stretch 
from the globule tip away from the main exciting star \citep{sugitani97}. 
This suggests a UV impact from the exciting star of 
IC1396 and a velocity pattern that may be explained by the collapsing inward 
gas motion \citep[e.g.,][]{leflo94}. 

Near-IR imaging revealed that several low-mass young stellar object (YSO) candidates were 
located at the globule tip facing the exciting star \citep{sugitani95}. 
These near-IR sources are nearly aligned along the globule axis \citep{sugitani97,sugitani99} 
and this object is considered to be a representative of the sequential 
star formation in BRCs. 
Some of them were identified as H$\alpha$ emission stars, 
which are most likely to be T Tauri stars \citep[hereafter OSP]{ogura02}, supporting 
the concept of sequential star formation. 
However, the details of the star formation were not clear from previous observations.
Since the masses and ages of the YSO candidates were still unknown,
we could not previously establish the sequence of star formation or its time scale. 

In order to estimate the masses and ages of the YSO candidates,
we have now confirmed their H$\alpha$ emissions with slitless spectroscopy and carried out 
additional optical and near-IR photometry.
Here we present our spectroscopic and photometric results and discuss sequential 
star formation in BRC37.
Additionally, we show our narrowband imaging of the $v$=1--0 S(1) H$_{2}$ line
to examine star formation activity of the YSO candidates.

Finally we discuss the candidate closest to the exciting stars, which seems from our photometric analysis 
to be a very low-mass object, i.e., a young brown dwarf candidate.  
Since sequential star formation from the direction of the exciting stars to the opposite side
is suggested for BRC37, this outermost candidate was likely formed by UV impact 
prior to the formation of other YSO candidates, and represents the first observational evidence
for brown dwarf formation as a consequence of UV impact.

\section{Observations}

\subsection{Optical observations}
Slit-less grism spectroscopy and $g$$\,'$ (482.5nm), $i$$\,'$ (767.2nm), and wide H$\alpha$ (651nm) 
imaging of BRC37 were conducted on 2006 November 19 UT and
$V$, and $I_{\rm C}$ imaging was done on 2007 August 7 UT with the Wide Field Grism Spectrograph 2  
\citep[WFGS2;][]{uehara04} mounted on the University of Hawaii 2.2-m telescope (UH88). 
The detector used was a Tektronix 2048 $\times$ 2048 CCD. 
The pixel scale is 0$\,''$.34 pixel$^{-1}$, yielding a field of view of 11.5$\,'$ $\times$ 11.5$\,'$. 
For the slitless spectroscopy, we used a wide H$\alpha$ filter (FWHM = 50nm) and a 300 line mm$^{-1}$ grism, 
providing a dispersion of 3.8 ${\rm \AA}$ pixel$^{-1}$.

Three 120 s exposures dithered by 10$\,''$ and single 60 s and 30 s exposures were taken at wide H$\alpha$, 
$g$$\,'$ and $i$$\,'$, respectively. 
For the slitless spectroscopy, we took three 300 s exposures dithered by 10$\,''$. 
The seeing was $\sim$0$\,''$.8 (FWHM) in  the wide H$\alpha$ image. 
For the photometric calibration the SDSS standard star BD +38$\degr$ 4955 \citep{smith02}
was observed at nearly the same airmass as the target
(difference $\lesssim$ 0.02). Twilight images were taken for flat-fielding.
The limiting magnitudes at 0.1 mag error level were $\sim$ 21.6 at $g$$\,'$ and $\sim$ 20.4 at $i$$\,'$. 

Three 60 s and three 30 s exposures dithered by 10$\,''$ were taken at the $V$ and $I$$_{\rm C}$ bands, respectively. 
The seeing was $\sim$ 1$\,''$.0 (FWHM) in the $V$ band. For the photometric calibration, 
the Landolt standard star \citep{Land92} was observed at nearly the same airmass as the target (difference $\lesssim$ 0.02). 
The limiting magnitudes at 0.1 mag error level were $\sim$ 21.7 at $V$ and $\sim$ 19.8 at $I$$_{\rm C}$.

\subsection{Near-infrared observations}

$J$ (1.25$\mu$m), $H$ (1.63$\mu$m), and $K$$_{\rm s}$ (2.14$\mu$m) band imaging of BRC37 was conducted 
on 2001 August 29 UT with SIRIUS \cite[the Simultaneous three-color InfraRed Imager for Unbiased Survey;][]{nag99,nag03} 
mounted on UH88. 
The SIRIUS camera, equipped with three 1024 $\times$ 1024 pixel HAWAII arrays, 
enables simultaneous observations in the $J$, $H$, and $K$$_{\rm s}$ bands. 
The pixel scale is 0$''$.288 pixel$^{-1}$, yielding a field of view of 4.9$\,'$ $\times$ 4.9$\,'$. 
The region of the near-infrared observation is shown as a square in Fig.\ref{fig1}.

We obtained thirty dithered frames with a 30 s exposure, resulting in a total exposure time of 900 s. 
The seeing was $\sim$ 1$\,''$.0 (FWHM) in the $K$$_{\rm s}$ band. 
The limiting magnitudes at 0.1 mag error level were $\sim$ 19.7 at $J$, $\sim$ 18.7 at $H$, 
and $\sim$ 18.0 at $K$$_{\rm s}$. 
We observed P9107 of the faint near-infrared standard star catalog \citep{per98} for the photometric calibration. 
Dome-flat frames were taken for flat-fielding.

Narrow-band imaging of  H$_{2}$ + continuum (2.12\micron) and continuum (2.26\micron)  was made on 1998 September 2 UT
with QUIRC \citep[the Quick Infrared Camera; ][]{hodapp} mounted on UH88. 
The pixel scale is 0$\,''$.189 pixel$^{-1}$, yielding a field of view of 3.2$'$ $\times$ 3.2$'$. 
We obtained ten frames of 150 s exposure dithered by 10$''$, resulting in a total exposure time of 1500 s 
for the H$_{2}$ line.
We also obtained ten dithered frames of 75 s exposure, resulting in a total exposure time of 750 s for the continuum.
Dome-flat frames were taken for flat-fielding.

\subsection{Data reductions}

We have applied the standard procedures for optical/NIR image reduction with IRAF (bias/dark subtraction, 
flat-fielding and averaging of dithered images).
The combined images of the wide H$\alpha$ and $JHK$$_{\rm s}$ bands are shown in Fig. \ref{fig1} and \ref{fig2}, respectively.  

We searched for H$\alpha$ emission stars in the combined image of slitless spectroscopy.   
For the detected candidates of the emission-line stars,
we have made 1D spectra with the aperture extraction package of IRAF 
and measured  their equivalent widths (EWs) with SPLOT. 

Stellar sources were initially identified by DAOFIND with a 4$\sigma$ detection threshold 
and all the  photometric measurements were made with the DAOPHOT package.
For the $V$ and $I$$_{\rm C}$ band photometry, we made rough linear
color-corrections from the observations of several standard stars \citep{Land92}, 
assuming average extinctions (0.10 mag airmass$^{-1}$
at the $V$ band and 0.08 mag airmass$^{-1}$ at the $I$$_{\rm C}$ band)
of Mauna Kea$\footnote{CFHT observatory Manual, Section 2\\
(http://www.cfht.hawaii.edu/Instruments/ObservatoryManual/CFHT$\_$Observatory$\_$Manual$\_$TOC.html)}$.
For the $JHK$$_{\rm s}$ magnitudes and colors, we made the color
conversion from the UH88/SIRIUS system to the CIT system
\citep{nag99}.

\section{Results}
\subsection{Selections of YSO candidates}
\subsubsection{Optical selection : H$\alpha$ emission stars}

\begin{figure}[ht]
\epsscale{1.0}
\plotone{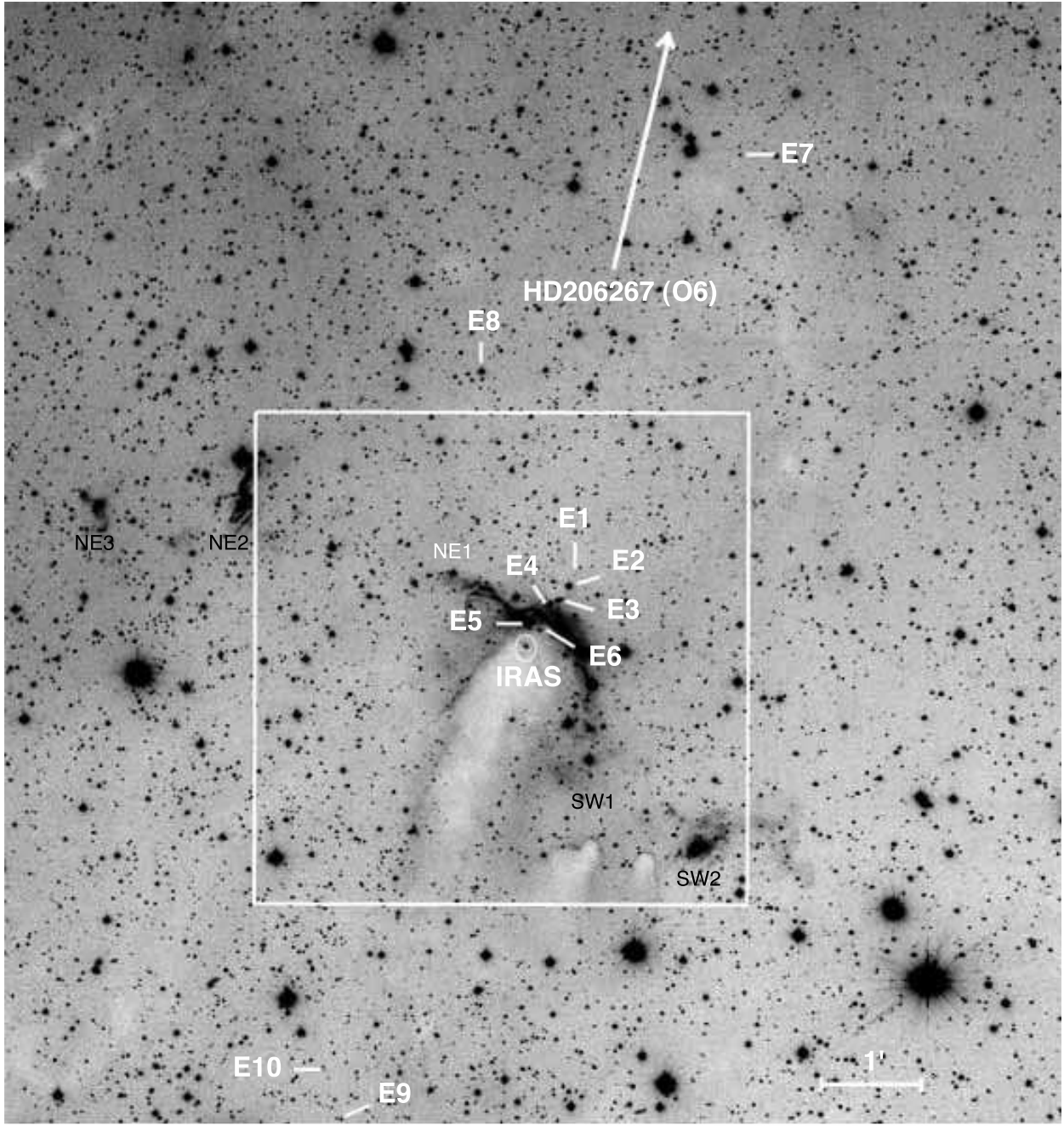}
\caption{\footnotesize {Image of BRC37 with wide H$\alpha$ filter. 
The position of IRAS 21388 + 5622 is shown with 
an error ellipse of position at the head part of the cometary globule. 
The size of the image is 11$\,'$.5 $\times$ 11$\,'$.5. 
The square is the region of the near-infrared observation. 
The H$\alpha$ emission stars are marked by E1-10. 
Known Herbig Haro objects are also indicated 
\citep[HH588 NE1, NE2, SW1, and SW2 and NE3, ][repsectively]{ogura02, froe05}. 
North is at the top and east to the left.}
}
\label{fig1}
\end{figure}

We searched for H$\alpha$ emission stars as YSO candidates.
We detected ten H$\alpha$ emission stars with slitless spectroscopy 
and refer to them as E1--E10 (Fig. \ref{fig1}).
Six of them (E1--E6) are located toward the tip of BRC37 (Fig. \ref{fig3} left). 
The EWs of the ten H$\alpha$ emission stars are listed with their coordinates in Table \ref{table1}.
E2--E5 and E7--E9 are candidates for classical T Tauri stars (CTTS) with EW $\ge$ 10 \AA.  
The continuum fluxes of E1, E6 and E10 are too faint to be measured, but these sources
are also good candidates for classical T Tauri stars 
since their H$\alpha$ emission lines are rather strong compared with their continua.
E6 has  [SII]$\lambda$6731, [NII]$\lambda$6584, and [OI]$\lambda$6300 emission lines (Fig. \ref{fig3} left). 

In the previous work \citep{ogura02}, the detection of eight H$\alpha$ 
emission stars was reported toward BRC37 (OSP 37-1 to -8; Fig. \ref{fig3} right).  

Six of these are confirmed here and cross referenced in Table \ref{table1}.
Additionally, as can be seen in Fig. \ref{fig3}, E4 is a strong H$\alpha$
emission source with weak continuum that is newly present in the 2006
observations. 
E4 is located just South of OSP 37-5. Its H$\alpha$ emission is clearly
seen in the slitless spectroscopic image from 2006
(Fig. \ref{fig3} left), whereas no counterpart is evident in the
1997 image (Fig. \ref{fig3} right).  Thus E4, as well as E7,
E9 and E10 (which lie outside the 1997 image), are newly detected
in these observations.  E7 and E8 are located north of BRC37, and E9
and E10 south of BRC37, being not associated with the BRC37 globule
(Fig. \ref{fig1}). 

However we can not confirm two H$\alpha$ sources from 1997, which
appear to have been misidentified in OSP.  OSP 37-5 is the continuum
source just to the North of E4. It was previously classified as a weak
H$\alpha$ source, but closer examination of both the 2006 and 1997
data reveals that its spectrum has a saw-toothed ripple, with a peak near
H$\alpha$. The narrow peak head of this ripple might be confused with
H$\alpha$ emission, but it is more likely an M star. 
This more careful analysis also reveals that OSP 37-4, to the East of
E2, was also misclassified as an H$\alpha$ source, probably due to
confusion with the overlapping continuum of E2. 

We noticed that the EWs in Table 5 of \cite{ogura02} are smaller than
those of this work, by about a factor of 3.8 on average.  Accordingly
we recalculated the EWs of the OSP stars from the 1997 data, using the
improved methodology applied in this work. We confirmed that this
discrepancy was due to the failure to scale from pixel to angstrom in
OSP, and not to any large variation of H$\alpha$ emission. 
Table \ref{table1} lists the EWs measured in this work for BRC37 in column 4, 
and those corrected from the 1997 data in the last column. 

The data presented in Table 5 of OSP were {\em not} scaled from pixel
to angstrom, and the data presented there for BRC13 are a repetition
of that for BRC12.  Accordingly we show here in an appendix as Table \ref{table4} 
a fully corrected version of Table 5 from OSP, recalculated with our
improved methodology, and properly scaled.  

\subsubsection{NIR selection : NIR excess sources}

We show the $J-H$ vs. $H-K$ color-color diagram in Fig. \ref{fig4}.
While E1 and E8 are below the T Tauri locus on the color-color diagram, 
E2--E7, E9, and E10 are located in the reddening zone of the T Tauri 
locus which indicates they are candidate classical T Tauri stars. 

In order to select sources having large near-infrared excess as YSOs
candidates, we searched for NIR sources that can be explained as
classical T Tauri stars with large NIR excess or those with
circumstellar envelope on the color-color diagram (Fig. \ref{fig4}).
We also examine E1--E10, which are identified as YSO candidates here. 

Since E7--E10 are outside of our NIR imaging area (Fig. \ref{fig1}),
we refered to the 2MASS data and additionally plotted E8 in
Fig. \ref{fig4}, after converting the 2MASS data to the CIT system
\citep{carpenter01}. 
We did not plot E7, E9 and E10; for E7 2MASS provides only upper
limits at $H$ and $K$$_{\rm s}$ and, while single sources are
cataloged/identified toward E9 and E10 in the 2MASS catalog/images,
our optical images show they have close stars with separations of
$\lesssim$ 2.0$''$, so their 2MASS colors are unreliable. 

\begin{figure}[ht]
\epsscale{1.0}
\plotone{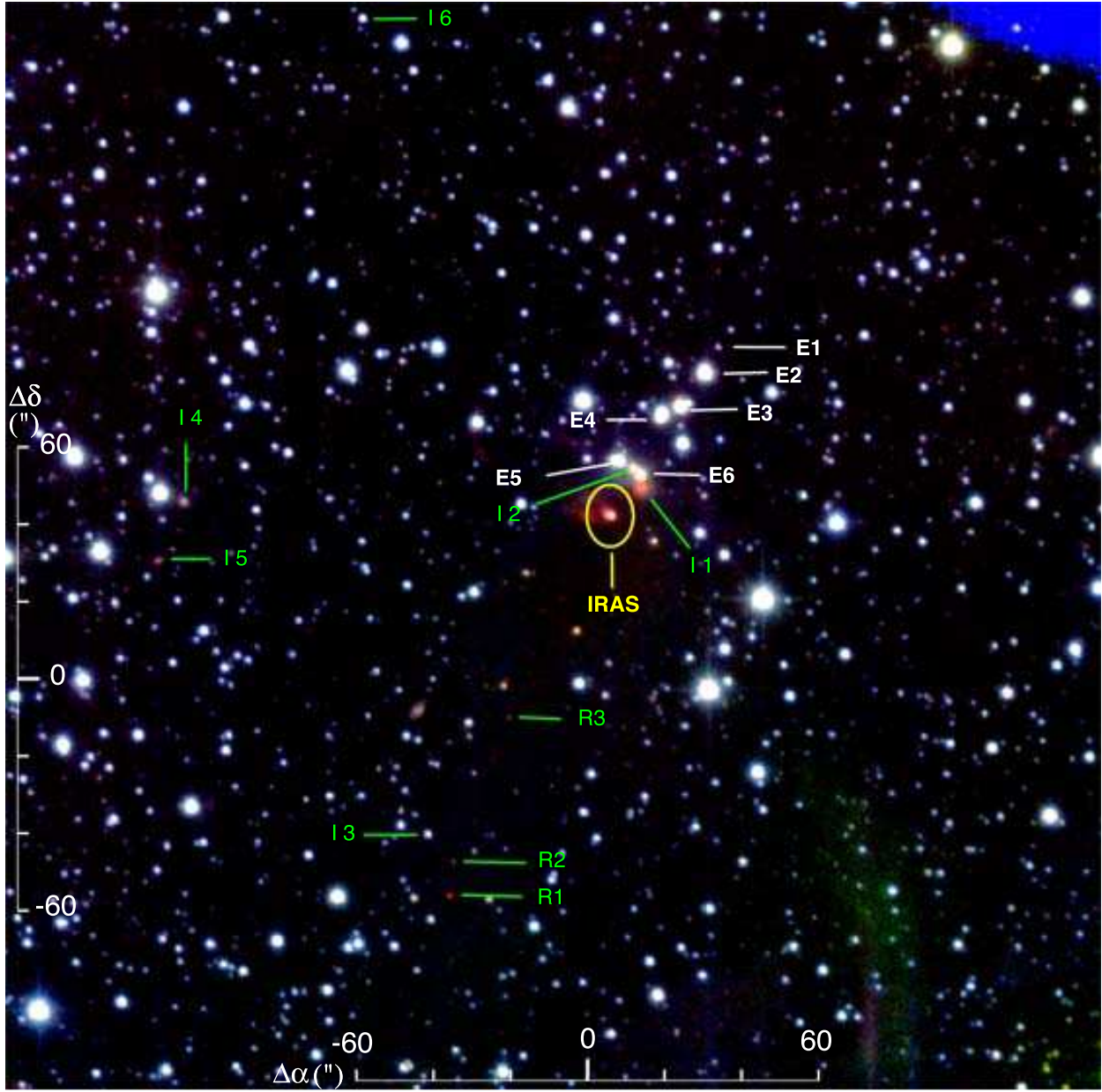}
\caption{Three-color composite image of BRC37 ($J$, blue; $H$, green; $K$$_{\rm s}$, red). 
The position of IRAS 21388 + 5622 is shown by an error ellipse of position.
The size of the image is $\sim$4.7$\,'$ $\times$ 4.7$\,'$.  North is at the top, and east is to the left. 
Offsets in arc seconds from the position of [$\alpha$(1950), $\delta$(1950)] = 
[21$^{\rm h}$38$^{\rm m}$54$^{\rm s}$.02, +56$^{\degr}$21$'$33$''$], 
which is the same reference position of \cite{duvert90}, 
are shown near the left and lower edges of the image.
The blue color region at the upper right corner is dead pixel region of the $J$ band array.
}
\label{fig2}
\end{figure}

\begin{figure}[ht]
\epsscale{1.0}
\plotone{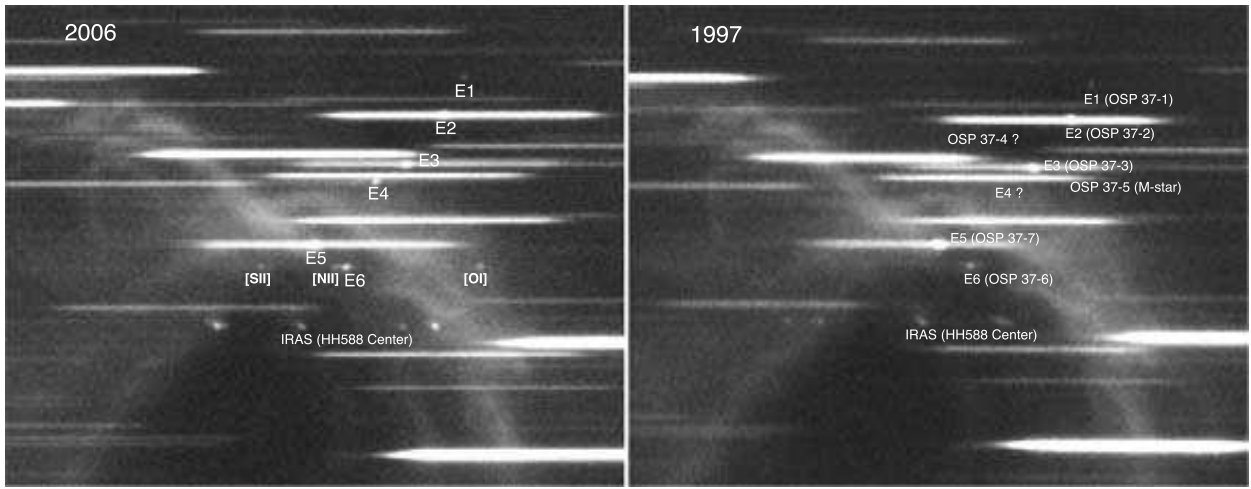}
\caption{Slit-less spectroscopic images of the BRC37 tip of this work on 2006 November 19 UT (left)  
and \cite{ogura02} on 1997 August 10 UT (right).  
H$\alpha$ emission of each H$\alpha$ emission star is marked by the ID of this work (E1--E6) 
and the OSP IDs are also shown (right).  
The left image was obtained with WFGS2 $+$ a wide H$\alpha$ filter (626.5--676.5 \AA), and the right image with WFGS $+$ 
a wide H$\alpha$ filter (6300--6750 \AA).   
Although the dispersions of these images are the same (3.8 ${\rm \AA}$ pixel$^{-1}$), 
the dispersion directions are different.
}
\label{fig3}
\end{figure}

\begin{figure}[ht]
\epsscale{0.6}
\plotone{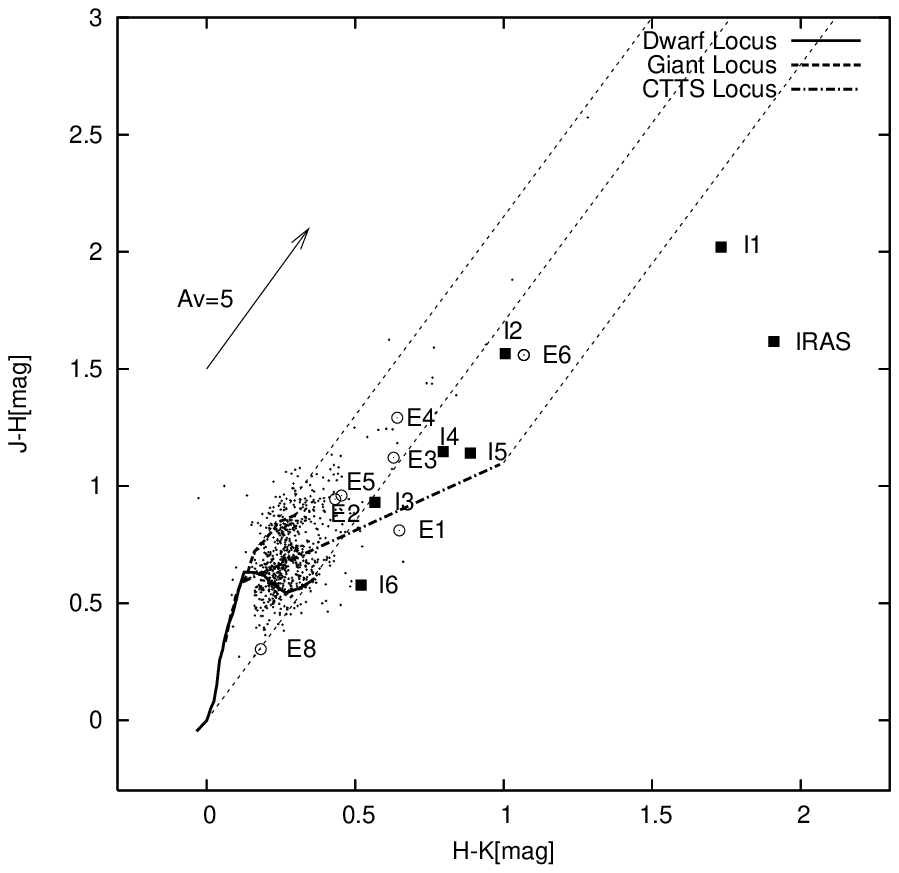}
\caption{$J - H$ vs. $H - K$ color-color diagram of BRC37. 
The solid and dashed curves are the loci of dwarfs and giants, respectively. 
The data for O9--M6 dwarf and G0--M7 giants are from \cite{bessell88}. 
The dash-dotted line is the unreddened CTTS locus of \cite{meyer97}.
Only the sources having photometric errors in each color of  $\le$0.1 mag are plotted.
}
\label{fig4}
\plotone{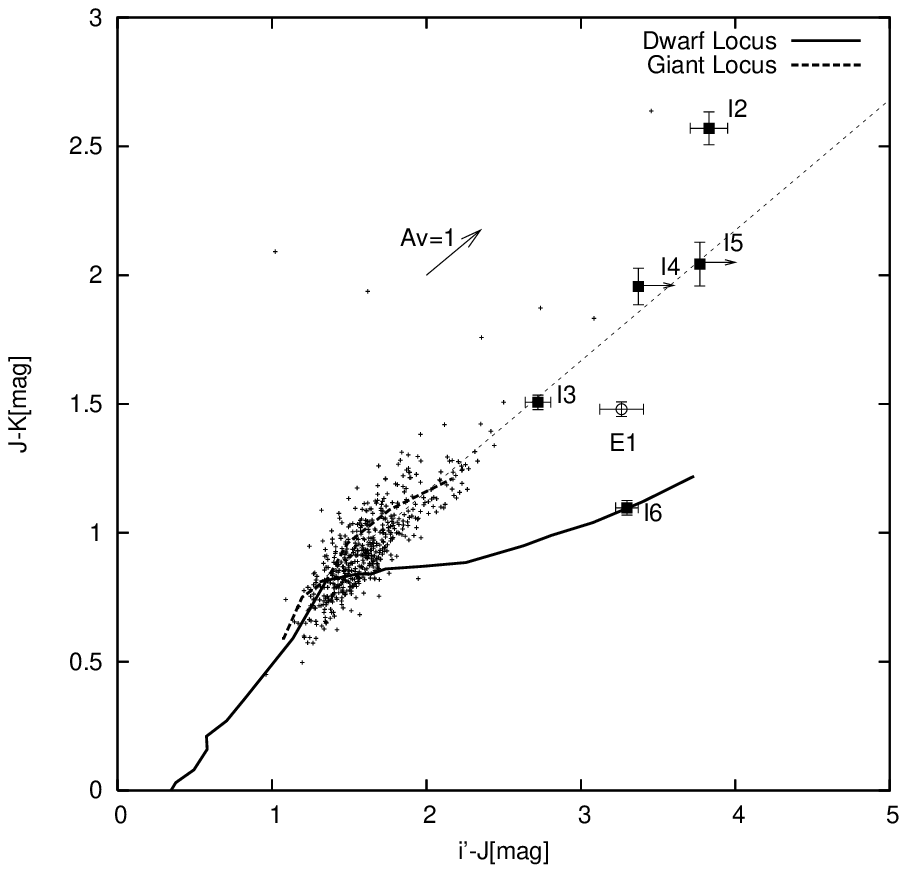}
\caption{$J - K$ vs. $i' - J$ color-color diagram of BRC37. 
The solid and dashed curves are the loci of dwarfs and giants, respectively. 
The dwarf locus for B8--K2 and K7--M7 is from \cite{bessell79,bessell91}.
The giant locus for G6--M4.5 is from \cite{bessell79}.
The thin dashed line shows the reddening line from a K7 star.
The sources having photometric errors in each color of  $\le$0.1 mag are plotted.
I2--I6 and E1 are plotted with errorbars.
}
\label{fig5}
\end{figure}

We included the sources located in the region surrounded by the
reddening band of main-sequence stars and giants, the reddening line
that originates from the truncated point of the T Tauri locus, and the
T Tauri locus of \cite{meyer97} with the color-colordiagram.  
We took photometric errors into account for this selection.
We excluded sources whose color errors could possibly make them those
without NIR excess.  We included those redward of the reddening line
from the truncated point.  This results in a total of seven NIR
sources selected here as YSO candidates. 
Two of them (the IRAS source and E6) have been identified as YSO candidates previously.
We refer to the other five NIR excess sources as I1--I5 (Fig. \ref{fig2}).
I1 and I2 are located toward the tip of BRC37, and I3 is located toward the tail of BRC37.
I4 and I5 are located east of BRC37. 

Several NIR-excess sources below the T Tauri locus (Fig. \ref{fig4})
are too faint ($J$ $\gtrsim$16) to be intermediate-mass YSOs
associated with this region, suggesting that they are background stars.
Two sources have the possibility that they are (young) substellar 
objects from the following analysis with another color-color diagram.  
One is E1 and the other is referred to here as I6 (Fig. \ref{fig4}). 

In order to eliminate the possibility that these candidates could be galaxies,
we use the $i'-J$ vs. $J-K$ color-color diagram (Fig. \ref{fig5}). 
We use the transform equation (Eq. 8 and Table 3) of \cite{jordi} to 
obtain the $i'$ magnitude of main-sequence stars and giants. 
We also compared this with our own transform equation obtained by 
observing several standard stars of \cite{Land92}.
We found the difference between $i'$ magnitudes calculated from these two 
methods to be small, $\lesssim$ 0.025 mag. in  a range of 
0$\lesssim R-I \lesssim$2.2, and therefore adopt the 
transformation of \cite{jordi}. 

Galaxies lie blueward of the reddening band of main sequence stars 
and giants \citep[e.g.,][]{nagshi03}.
From this diagram, I2 is judged to be a galaxy. 
I3 is located near the boundary of the reddening band.
Our slitless spectroscopic image shows no H$\alpha$ emission for I3, 
suggesting that it is not a T Tauri star associated with the tail of BRC37.  
Taking into account its errors of color magnitude, it is also likely to be a galaxy. 
I4 and I5 are too faint to be measured in the $i'$ band image and only 3$\sigma$ upper 
limits is given. 
Although I4 can not be classified, the appearance of I5 on the $JHK$$_{s}$ images 
suggests that I5 is a galaxy.

The position of I1 is  [$\alpha$(2000), $\delta$(2000)] = 
[21$^{\rm h}$40$^{\rm m}$27$^{\rm s}$.94, +56$^{\degr}$36$'$02$''$.38].
The color-color diagram can not be used for I1 since there is no detection 
at $i'$. 
However I1 has a large NIR excess and a NIR nebulosity (Fig. \ref{fig2}), 
and is most likely to be an embedded YSO inside BRC37. 

Only E1 and I6 have the colors of late M-dwarfs or reddened late M-dwarf 
on the color-color diagram, 
leaving the possibility that these are (young) substellar objects. 

\subsubsection{NIR Selection : More Embedded Sources ?}

In order to select embedded Class I candidates in BRC37 and its neighboring clouds  
(two small clouds south-west of BRC37, Fig. \ref{fig2}), 
we examined NIR sources that are detectable both at $H$ and $K$$_{\rm s}$, but not at $J$,
and found seven such sources by this method. 
We further examined their association with molecular cloud cores. 
\cite{baba06} reported that most of protostar (Class I) candidates are associated with the C$^{18}$O clumps distributed 
throughout the cloud C of the Vela Molecular Ridge by their deep NIR imaging. 
Here we consider the sources located toward the centers of cloud cores to be possible Class I candidates. 

Following these selection criteria, we list three ``red'' NIR sources named R1--R3 (Fig. \ref{fig2}). 
R1 has an $H-K$ color magnitude of $\sim$2.4, 
which suggests a possibility that R1 is a Class I source embedded in the tail of BRC37.
The position of R1 corresponds to the peak located at the (45$''$, --56$''$) offset position 
in  the $^{13}$CO ($J$=1--0) channel map of $V_{\rm LSR}$ = 1.25 km s$^{-1}$ \citep[Fig. 4b of ][]{duvert90}. 
R2 is $\sim$0.15$'$ north of R1, near the (45$''$, --56$''$) $^{13}$CO peak, 
and R3 is $\sim$0.76$'$ north-northwest of R1, toward the center of the (20$''$, --8$''$)  
$^{13}$CO peak. 
R2 and R3 have $H-K$ color magnitudes of $\sim$1.1--1.2, which are similar to those of reddened T Tauri stars 
with near-IR excess (e.g., E6 in Fig. \ref{fig4}), 
but \cite{baba06} showed that some class I sources appear in the $J-H$ vs. $H-K$ diagram 
in the region of reddened T Tauri stars having NIR excess. 
Because R1--R3 are located toward CO intensity peaks, i.e., molecular clumps, 
all of them are possible Class I candidates.

\subsubsection{H$_{2}$ ($v$=1--0, S(1)) emission-line}

\begin{figure}
\epsscale{1.1}
\plottwo{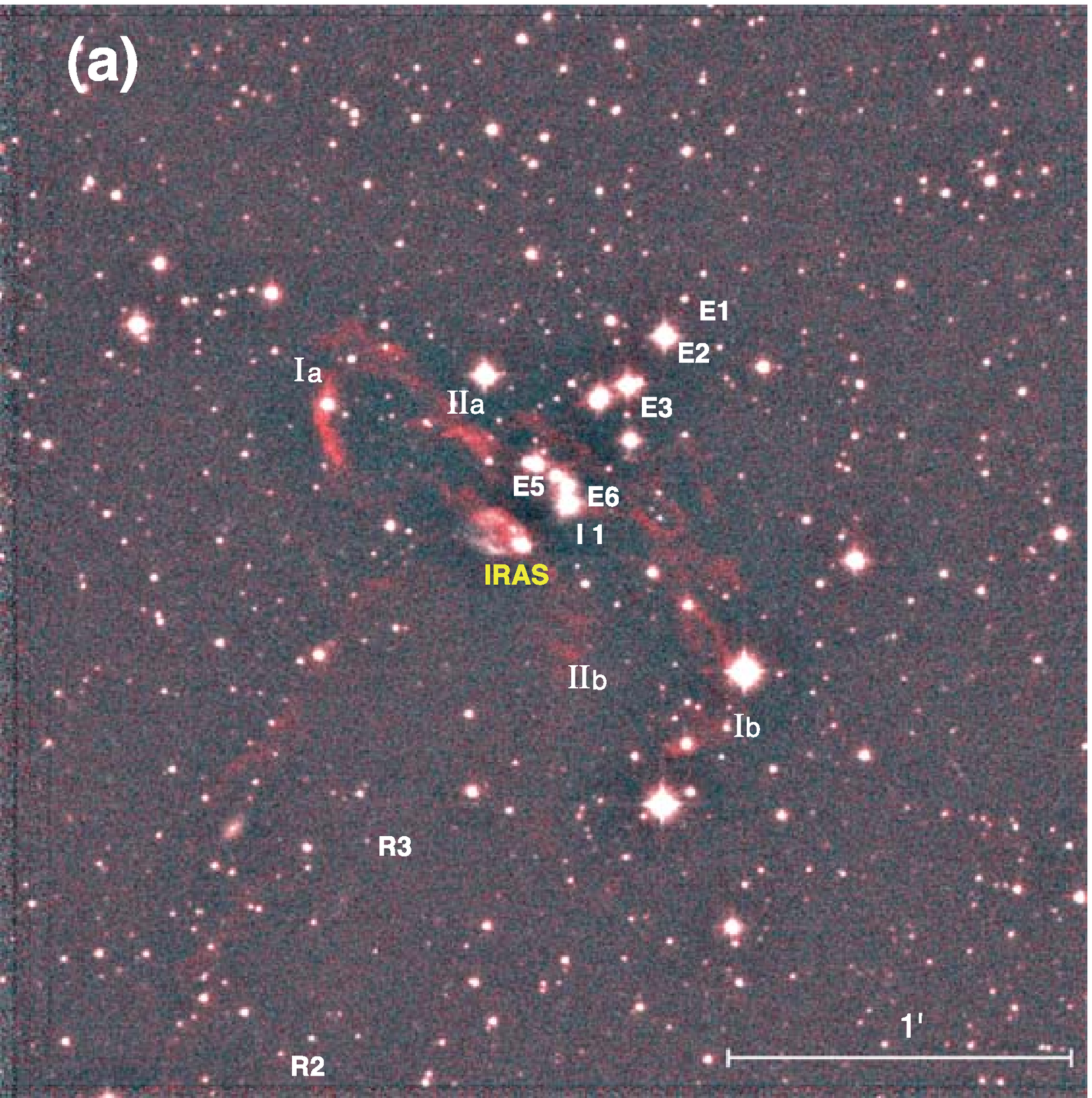}{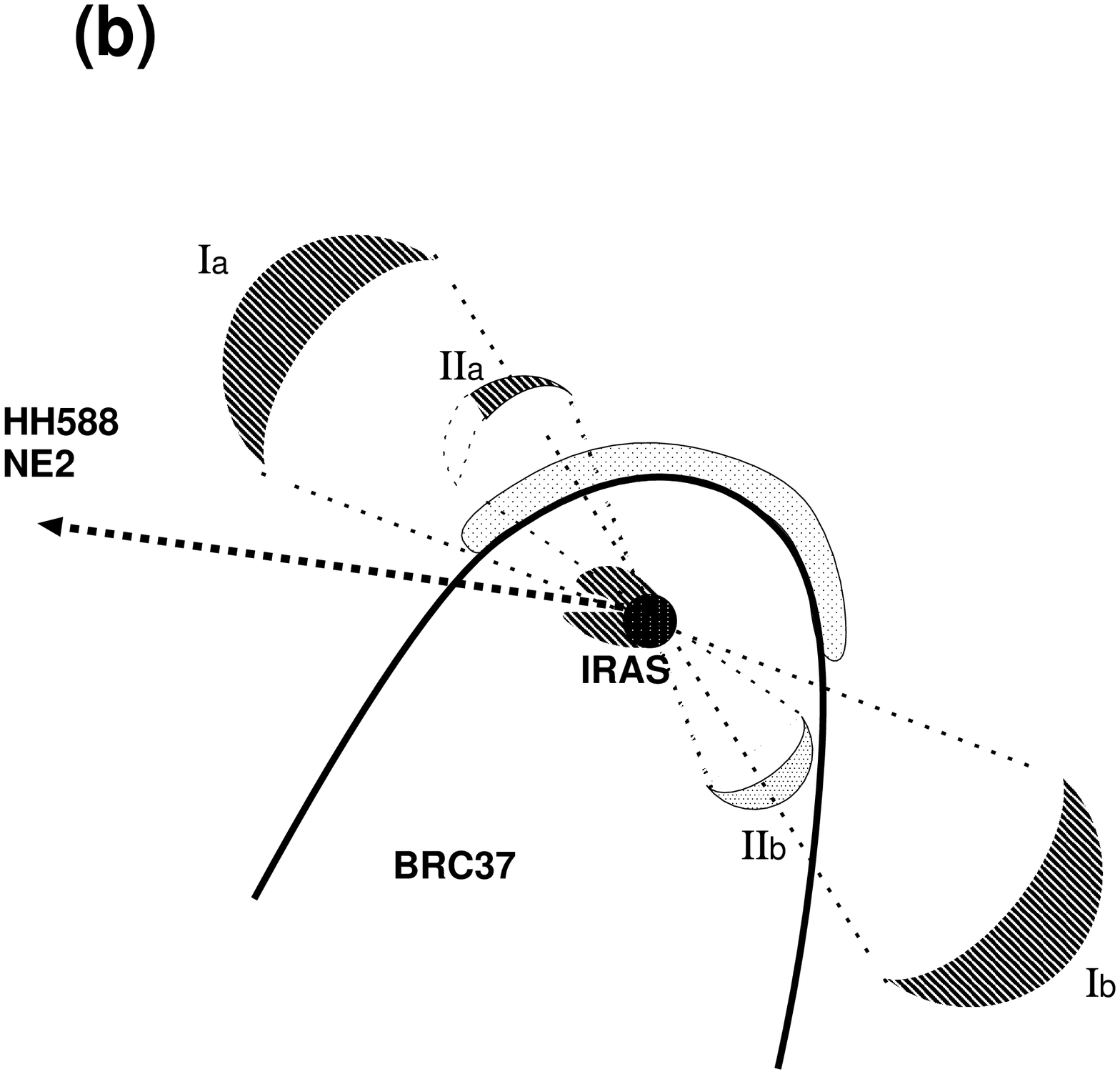}
\caption{(a) False color image (cont., blue; cont., green; H$_{2}$+cont., red)
of the BRC37.  The size of image is $\sim$3.2$'$ $\times$ 3.2$'$.  
(b) The schematic drawing of H$_{2}$ outflow structures. 
 }
\label{fig6}
\end{figure}

In order to examine star formation activities of the YSO candidates,
we searched for H$_{2}$ emission around them.
We show a false color image of H$_{2}$ emission and continuum in Fig. $\ref{fig6}a$.

Toward the immediate east-vicinity of the IRAS source, two lobes that extend from the IRAS source are shown. 
They probably correspond to the walls at the root of the cavity that was created by the outflow pointing
to the NE2 knot of HH588 (P.A. $\sim$60$\degr$  from the IRAS source).
A bow structure (Ia) can be identified $\sim$0$'$.7 northeast of 
the IRAS source (P.A. $\sim$54$\degr$ from the IRAS source). 
This probably corresponds to the NE1 A and B knots of HH588.
A counterpart (Ib) of this bow structure is also identified $\sim$0$'$.7 southwest of 
the IRAS source (P.A. $\sim$220$\degr$ from the IRAS source).
In addition, a partial, compact bow structure (IIa) is also identified $\sim$0$'$.4 north-northeast of 
the IRAS source (P.A. $\sim$32$\degr$ from the IRAS source). 
A counterpart (IIb) of this compact bow structure is faintly identified  $\sim$0$'$.4 south-southwest of  
the IRAS source (P.A. $\sim$206$\degr$ from the IRAS source),
while it is invisible in the optical.
In Fig. $\ref{fig6}$b, we show the schematic drawing of these H$_{2}$ structures, 
which are considered to be geometrically related
to the outflow activities of the IRAS source and thus are likely to be shock-excited.

\subsection{Examination of YSO Candidates}

Table. \ref{table2} shows the properties of the YSO candidates, which are derived from our photometry.

We examined the photometric variability at $i'$ and $g'$ for our detected H$\alpha$ emission stars 
by comparing another WFGS2 observation that was conducted on 2006 September 22. 
Although these stars showed small differences between the two epochs ($\lesssim$0.6 mag), 
the variability is not clear due to their faintness and 3$\sigma$ errors of $\sim$0.1--0.3 mag in relative photometry. 

\begin{figure}[ht]
\epsscale{1.0}
\plottwo{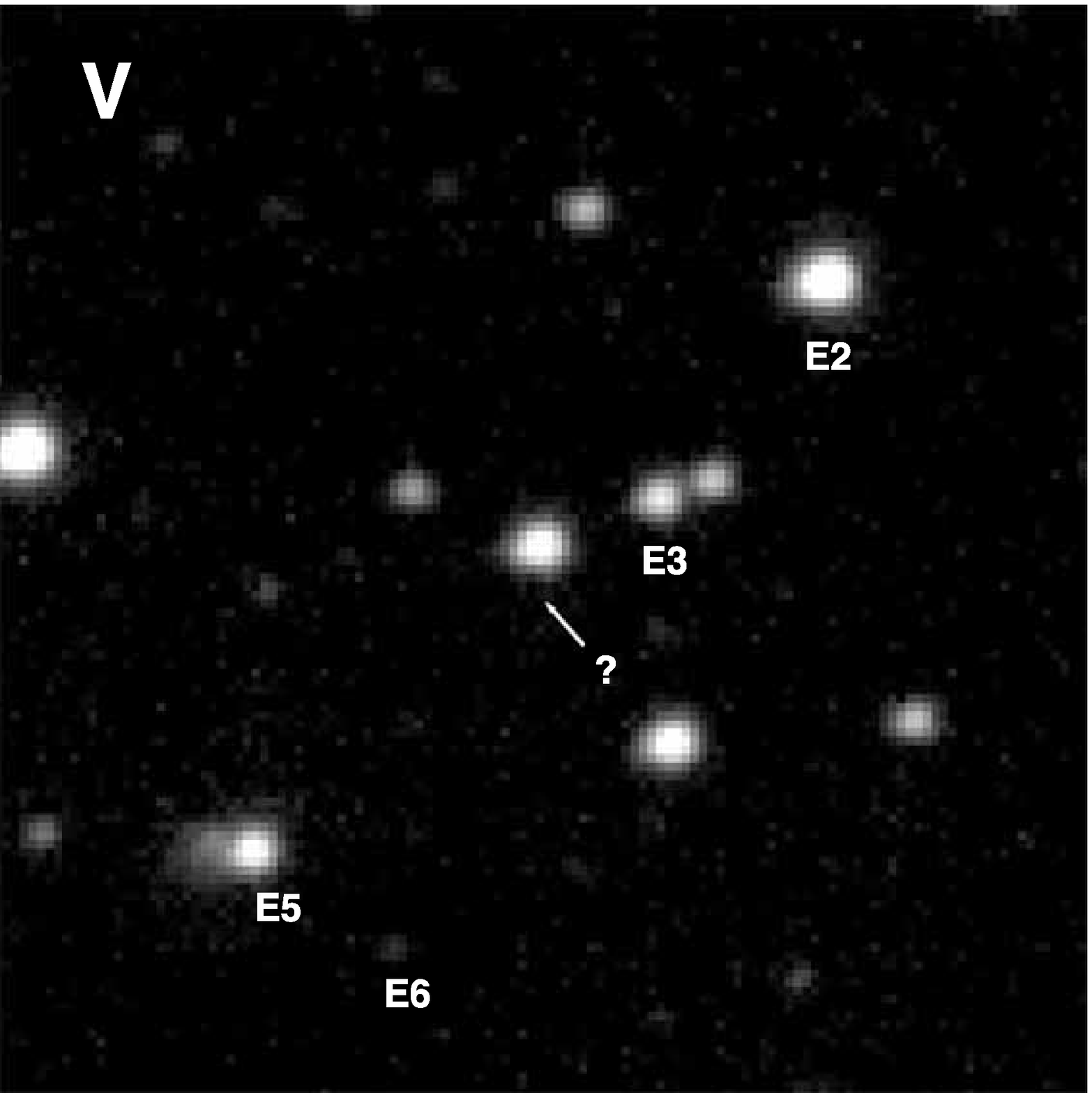}{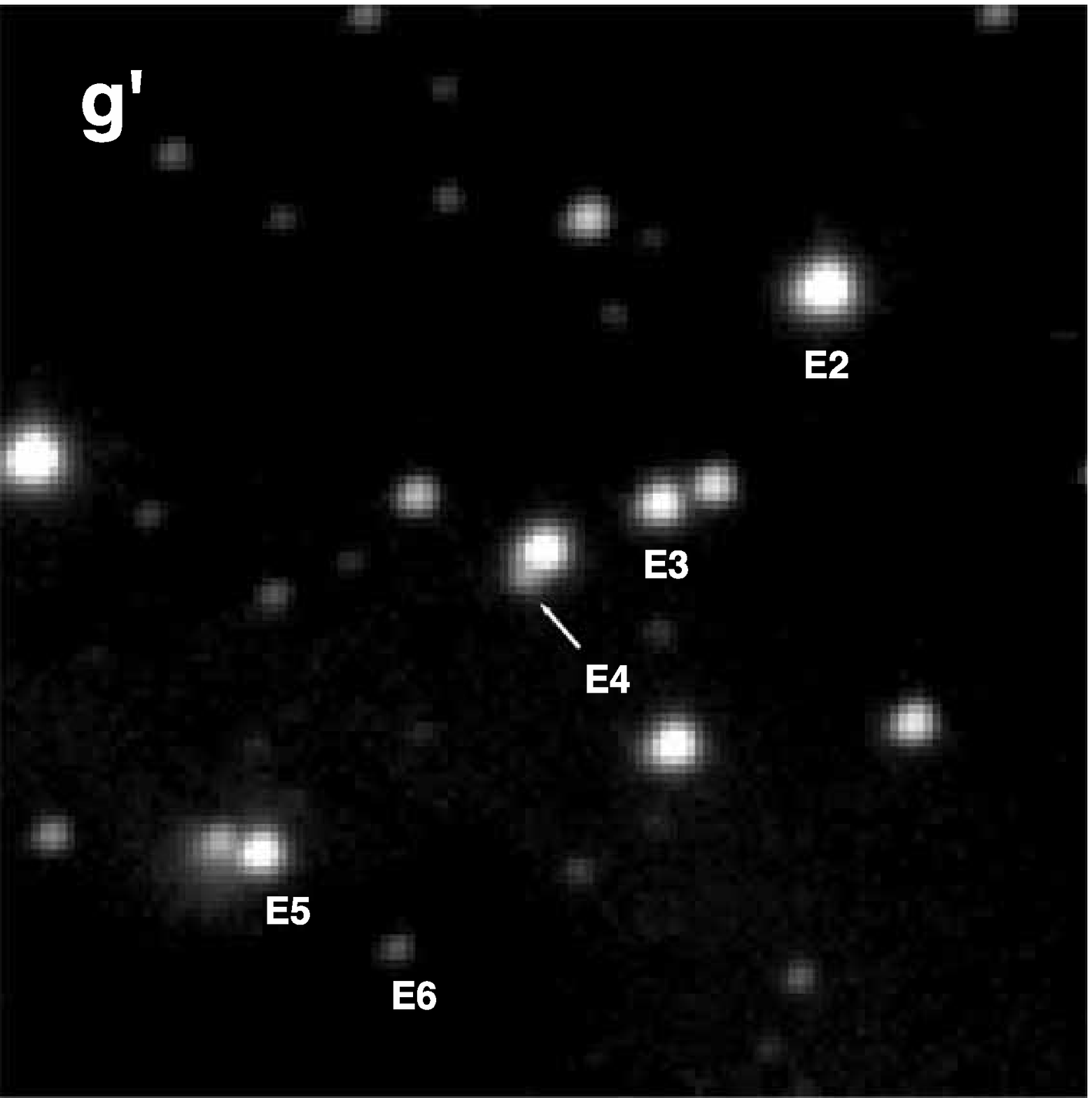}
\caption{Images of E2--E6 at $V$ (left) and $g'$ (right) bands in logarithmic scale.
E4 is clearly identified in the $g'$ band image, but not in the $V$ band image. 
The $V$ band image was taken on 2007 August 7 UT
and the $g'$ band image on 2006 November 19 UT.
North is at the top, and east is to the left. 
The size of each image is $\sim$42$\,''$ $\times$ 42$\,''$. 
}
\label{fig7}
\end{figure}

We further compared the $g'$ image with the $V$ image
(Fig. \ref{fig7}).  The intensities of E2, E3, E5 and E6 in the $g'$
and $V$ images are similar, whereas E4 can not be identified on the
$V$ image (Fig. \ref{fig7} left). 
Precise measurement of this variation is difficult due to
contamination from a neighboring brighter star, but our rough estimate
indicates that the magnitude difference is $\sim$ 3 mag. or
larger. 
A similar variation is also seen between the $I$$_{\rm C}$ and $i$$'$
images.  Thus E4 may be a background, cataclysmic variable star with
H$\alpha$ emission, although we cannot eliminate the possibility of
occultation due to the circumstellar material of E4 (e.g., a
precession disk, inhomogeneous disk structure, etc.).  The variation
of E4 is very rare, and follow-up observations are required to monitor
it. 

We also examined the variability of H$\alpha$ emission for the
H$\alpha$ emission stars.  The variability of the EWs for E2, E3, E5,
and E8 is at several ten percent level over the 10 yr period (Table
\ref{table1}), and seems to be natural for CTTSs.  As was mentioned in
Section 3.1.1, the H$\alpha$ emission of E4 is identified in the
spectroscopic image of this work, but not in that of the previous work
from 1997.  This again indicates very large variability of H$\alpha$
emission for E4. 

For E6, [SII]$\lambda$6731, [NII]$\lambda$6584, and [OI]$\lambda$6300
emission lines are seen in the spectroscopic image of this work, but
not in that of the previous work (Fig. \ref{fig3} right).  This
variation may indicate outflow variability of E6 or a clearance of its
circumstellar material. 

\begin{figure}[ht]
\epsscale{0.6}
\plotone{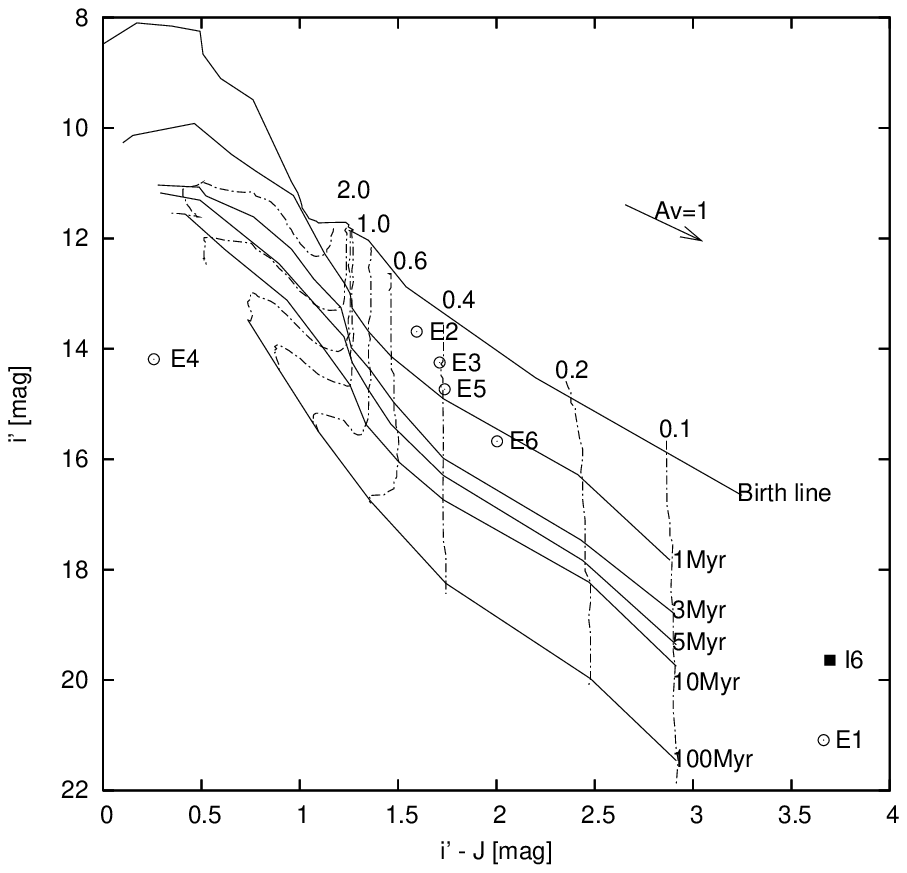}
\caption{Extinction-corrected $i'$ vs. $i' - J$ color-magnitude diagram of YSO candidates. 
H$\alpha$ emission stars are  shown by open circles. 
The birthline and the 1, 3, 5, 10, and 100 Myr isochrones of \cite{palla99} are shown with 
the evolutionary tracks for masses from 0.1 to 2.0 M$_\sun$, 
at an assumed distance of 750pc. 
Extinction is not corrected for E1 and I6.
}
\label{fig8}
\epsscale{0.6}
\plotone{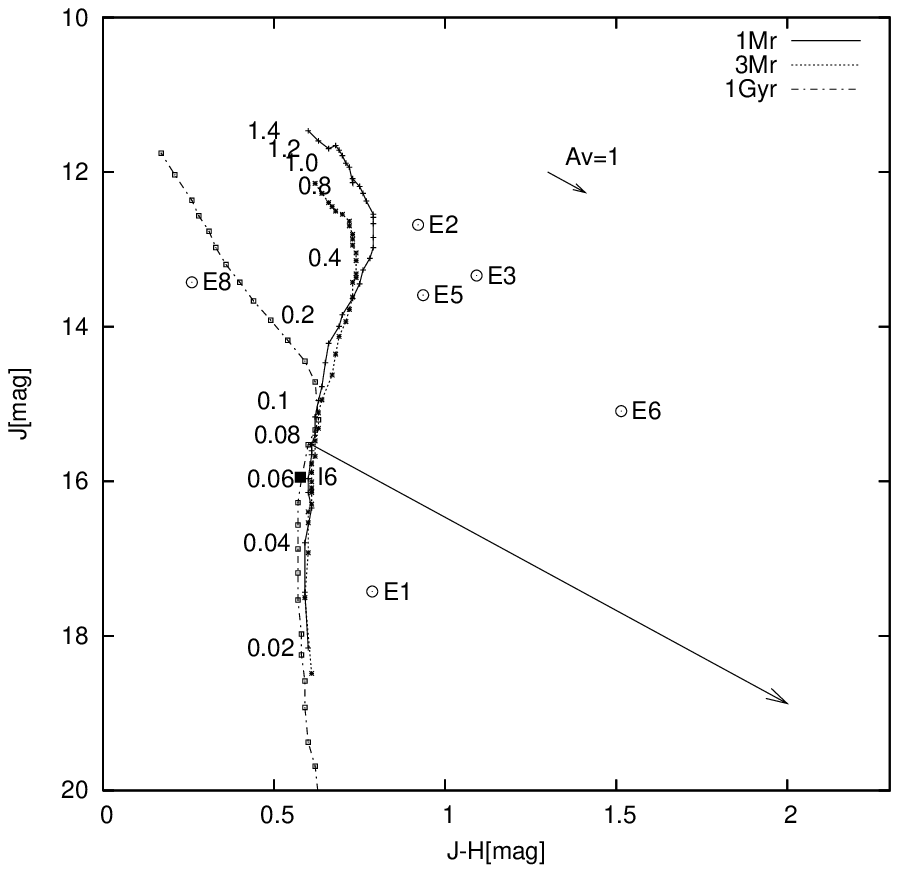}
\caption{$J$ vs. $J - H$ color-magnitude diagram of YSO candidates. 
H$\alpha$ emission stars are shown as open circles. 
The 1 Myr,  3 Myr and 1 Gyr isochrones of the NextGen model \citep{baraffe98} 
at an assumed distance of 750pc are shown. 
The reddening line from a 0.075 M$_\sun$ star of the 1 Myr locus is indicated (by arrow).
The numbers near the 1 Myr locus indicate star masses of the 1 Myr locus.
}
\label{fig9}
\end{figure}

We can estimate the masses and ages of the five H$\alpha$ emission stars (E1--E3, E5, and E6) 
by assuming that they are classical T Tauri stars associated with IC1396 ($d$=750pc). 
We made an extinction corrected color-magnitude diagram 
($i'$ vs. $i'-J$ diagram, Fig. \ref{fig8}) with the evolutionary tracks and isochrones of \cite{palla99}. 
Additionally, we used $A_{i'}$=0.652$A_{V}$ derived from \cite{cardelli89} for extinction correction.
We adopted the T Tauri locus of \cite{meyer97} as  the intrinsic colors of these emission stars, except for E1, 
and evaluate their extinctions on the $J-H$ vs. $H-K$ diagram by using the reddening law of \cite{cohen}.
By this method we calculate that E2, E3, E5, and E6 have ages of $\sim$ 1 Myr and masses of $\sim$0.3--0.5 M$_\sun$.  
We also constructed $I$$_{\rm C}$ vs. $I$$_{\rm C}$$-J$ diagram and found slightly smaller masses ($\sim$ 0.2--0.4 M$_\sun$).
Since these masses and ages are typical for T Tauri stars, we conclude that E2, E3, E5, and E6 
are very likely to be  T Tauri stars.  

We note that E4 is located below the 10 Myr isochrone on the $i'$
vs. $i'-J$ diagram, suggesting that it is not a T Tauri star
associated with IC1396. 

It is difficult to evaluate the mass and age of E1 from the $i'$ vs. $i'-J$ 
diagram with the evolutionary tracks and isochrones of \cite{palla99};
its magnitude is very faint and its extinction cannot be estimated from its color-magnitude location, 
which is below the T Tauri locus on the $J-H$ vs. $H-K$ diagram (Fig. \ref{fig4}). 
Instead, we constructed $J$ vs. $J-H$ color magnitude diagram (Fig. \ref{fig9}), 
adopting the isochrones of the NextGen model \citep{baraffe98}.
The reddening line, which originates from a 0.075 M$_\sun$ star of the 1.0 Myr locus, 
indicates the boundary between H-burning stars and brown dwarfs.
In this diagram, E1 lies below this reddening line, suggesting that it is a young brown dwarf with H$\alpha$ emission.
Even if the 3 Myr locus is adopted, E1 again falls below the reddening line from a 0.075 M$_\sun$ star. 

We estimated $A_{V}$ $\sim$ 2.0 for E1 with the 1 Myr locus in Fig. \ref{fig9}. 
Even with the 3Myr locus, the estimated value is the same.
However, E1 could have NIR excess because it has a disk with strong H$\alpha$ emission. 
Recently, it was reported that young brown dwarfs have circumstellar disks, i.e., 
strong H$\alpha$ emission due to mass accretion (e.g., Luhman et al. 2007).
If E1 actually has NIR excess, this extinction becomes an upper limit.
This upper limit is similar to $A_{V}$ of E2, which is located next to E1 
and a little closer to BRC37.

\begin{figure}[ht]
\epsscale{0.5}
\plotone{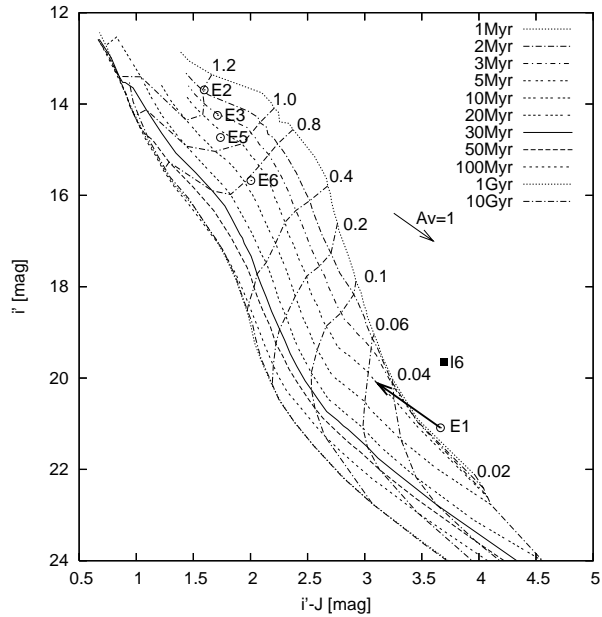}
\caption{Extinction-corrected $i'$ vs. $i' - J$ diagram of YSO candidates. 
The isochrones of the NextGen model \citep{baraffe98} are shown with the evolutionary tracks
for masses from 0.02 to 1.2 M$_\sun$ at an assumed distance of 750pc.  
}
\label{fig10}
\end{figure}

We show the $i'$ vs. $i'-J$ diagram with the isochrones of the NextGen model 
\citep{baraffe98} in Fig. \ref{fig10}.
Without any extinction correction, the location of E1 in this diagram indicates that it is a young brown dwarf. 
If we adopt the upper limit value, the mass and age are estimated to be $\sim$0.06 M$_\sun$ and $\sim$5.0 Myr, respectively.
If E1 has extinction smaller than 2.0 mag., its mass and age become smaller.
The derived age of $\sim$5.0 Myr for E1 (assuming an upper limit for reddening) is somewhat larger than the estimated ages of  
E2, E3, E5, and E6 in Fig. \ref{fig8}.  

With the Nextgen model, the ages and masses of E2, E3, E5, and E6 are estimated to be   
$\sim$2--8Myr and $\sim$0.8--1.2 M$_\sun$, respectively. 
The values are somewhat larger than those from the model of \cite{palla99}.
\cite{luhman03} used the evolutionary model of \cite{palla99} for M/M$_\sun$ $\geq$ 1 and
the NextGen model for M/M$_\sun$ $\leq$ 1.
Following \cite{luhman03}, we use the model of \cite{palla99} for E2, E3, E5, and E6 and
we use the NextGen model for E1.
From these comparisons, we conclude that E1 is most likely to be a young brown dwarf with an age of a few Myr. 

For I6, no H$\alpha$ emission was detected.
This suggests  that I6 has no circumstellar disk, and therefore is not so young.
If I6 is located at the distance of IC1396 (750pc), it would have a very small extinction (Fig. \ref{fig9}),
and then would be plotted far above the isochrone of 1 Myr on Fig. \ref{fig10}.
Specifically, if we assume a young age of $\sim$10Myr for I6, we can estimate its 
distance and mass to be $\sim$240pc and $\sim$0.03M$_\sun$, respectively.
Alternatively if I6 is very old ($\sim$1--10Gyr), a distance and mass of $\sim$100pc and $\sim$0.085M$_\sun$ 
are indicated, i.e., a late M-dwarf. 
In either case we conclude that I6 is a foreground source. 
 
In order to further examine E7--E10, we constructed the $i'$ vs. $g'-i'$ 
and $I$$_{\rm C}$ vs. $V-I$$_{\rm C}$
diagrams with the isochrones of \cite{palla99}. 
E7, E8 and E10 are located below the isochrone of 10 Myr and 
therefore are likely to be background stars.
From the $J-H$ and $H-K$ colors, E8 is likely to be a Herbig Ae/Be star or classical Be star.
If extinction is large, E9 may be a background star.
But if extinction is small, there is a possibility that E9 is a 
YSO associated with IC1396 with an age of $\sim$ 3--10 Myr. 
We note that E9 shows variability:
Its $i'$ magnitude was measured at 17.7 on 2006 November 19 UT and its 
$I$$_{\rm C}$ magnitude  was 16.1 on 2007 August 7 UT.
We can expect an i-I difference for this red object of $\leq$~0.8 mag from the differences in photometric systems,
$\footnote{\cite{lup05}; http://www.sdss.org/dr5/algorithms/sdssUBVRITransform.html\#Lupton2005}$
so the large measured difference ($\sim$ 1.6 mag.) indicates substantial brightening over this period. 

In Table\ref{table2}, we summarize the YSO identifications in this work. 
Objects in the  upper part of this table are the YSO candidates associated with the tip of BRC37 while
those in the middle part are the YSO candidates located toward the tail part of BRC37.
Objects in the lower part are considered to be background/foreground sources, 
except E9 whose uncertain status is discussed above. 

\section{Discussion}

\begin{figure}
\epsscale{1.0}
\plottwo{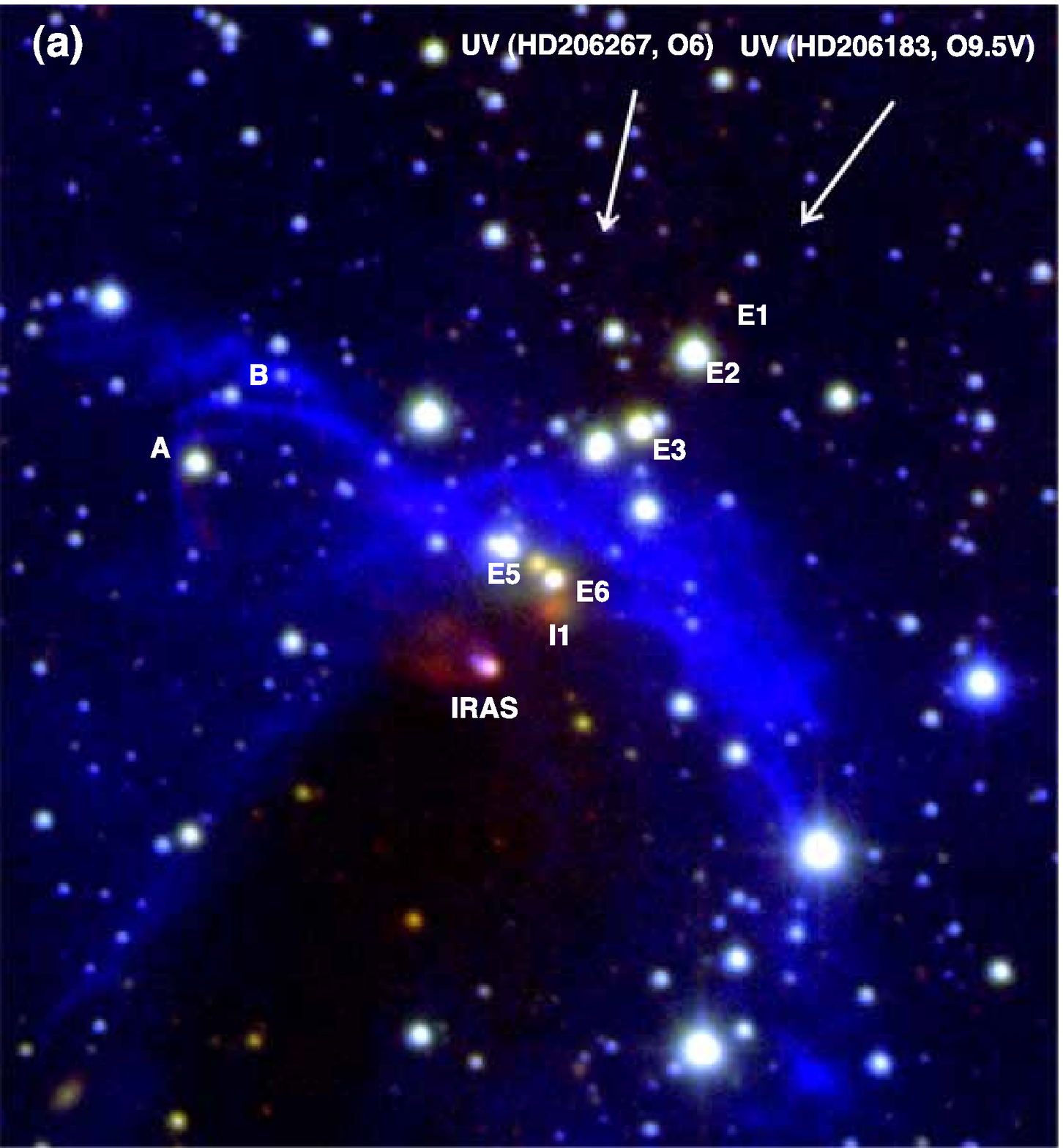}{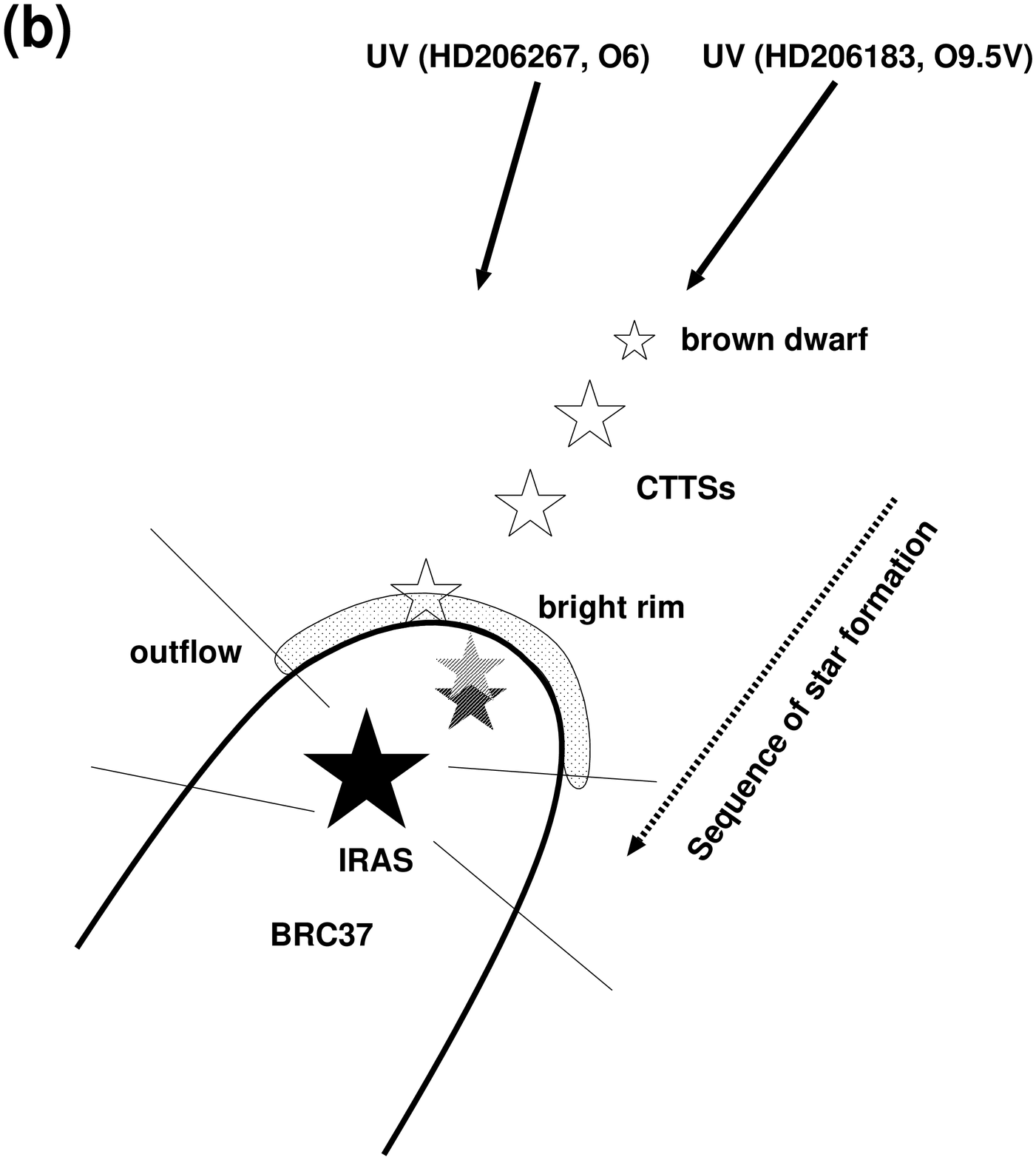}
\caption{(a) Three composite image (wide H$\alpha$, blue; $H$, green; $K_{\rm s}$, red) 
of the closeup of the tip of BRC37 and (b) the schematic drawing.
The knots A and B of HH588 NE1 are marked by labels A and B.
The size of image is $\sim$ 2.0$'$ $\times$ 2.0$'$.
}
\label{fig11}
\end{figure}

In this section, we discuss only the objects in the upper and middle parts of Table$\ref{table2}$ 
as YSO candidates of  BRC37.
In Fig. \ref{fig11}$a$, a closeup view toward the tip is shown with the labels 
of these YSOs including the IRAS source. 

\subsection{Sequential star formation in BRC37}

The IRAS source is still driving the H$_{2}$ bow shocks and HH objects, and 
is considered to be a protostar-like object having a considerably young age, 
whereas the dynamical age of the molecular outflow was reported to be $\sim$ 0.3 Myr \citep{duvert90}.
On the other hand, the ages of E1--E3, E5, and E6, which are located at the globule tip 
facing the exciting star(s), are estimated to be $\sim$ 1.0 Myr.
These suggest that star formation has been taking place from the outside of the cloud toward the IRAS source. 

In more detail, it seems that E6 and I1 are younger than the objects on the closer side of the exciting star(s), 
and are older than the IRAS source.
They are located between E1--E3, and E5 and the IRAS source.
While E1--E3, and E5 are outside the molecular cloud, E6 and I1 are embedded in 
the $^{13}$CO molecular gas \citep[see Fig. 2 of ][]{sugitani97}, 
and have considerably larger NIR excess than those outside sources.
E6 has the [SII], [NII], and [OI]  lines, while the other emission stars (E1--E3 and E5) do not show 
these lines, suggesting outflow activity and a younger age for E6. 
I1 has a NIR reflection nebula, which suggests association with a circumstellar envelope.
However, since I1 and E6 do not show strong outflow activities like the IRAS source, 
they are older than the IRAS source. 
These findings indicate sequential star formation in order of E1--E3 then E5, E6 and I1, then the IRAS source 
from outside the cloud to the central source itself (Fig. \ref{fig11}b). 

\subsection{Direction of exciting sources} 

We examined the direction from the main exciting star to the IRAS source (UV incident angle)
 and the axis of  the tip part of BRC37.
We determined the axis/elongation direction by eye using the $^{13}$CO maps of \cite{sugitani97}.
This elongation direction of $^{13}$CO gas at its tip part (P.A. $\sim$165$\degr$) is nearly parallel
to the direction of the UV radiation from the main exciting star to the IRAS source (P.A.=166.6$\degr$).
This suggests that the tip structure might have been strongly affected 
by the UV radiation as was discussed in \cite{sugitani97}.
The axis direction of the head part, which is a region of $\sim$1$'$.5 long and includes
the tip part, is estimated to have P.A. $\sim$147$\degr$ (Fig. \ref{fig1}) and slightly differs from
the two directions above by $\sim$20$\degr$.

Interestingly, the YSOs are almost aligned along a straight line (Fig. \ref{fig11}$a$), 
whose direction (P.A. $\sim$147$\degr$) differs from that of the UV radiation 
from the main exciting star by $\sim$20$\degr$, 
but agrees with the axis direction of the head part (Table \ref{table3}). 
Therefore, we searched for other exciting stars along the straight line from the YSOs 
toward the central region of IC1396, and 
found two candidates, which are members of Trumpler 37 \citep{garrison76},
on a straight line of P.A. $\sim$143$\degr$. 
One is HD205794 (B0.5V) and another is HD206183 (O9.5V).
The former is located toward the central region near the rim Ab cloud \citep{wei96} and
the projected distance is $\sim$14pc, while the latter is located at a closer position
with a projected distance of $\sim$6pc.
Since the latter has an earlier type and is closer to BRC37, it is  possibly another exciting star of BRC37.
However, the flux of ionizing photons from the main O6 exciting star is estimated to be $\sim$5 times larger 
than that from HD206183 (O9.5V) at the position of BRC37 and the impact from the main exciting star 
seems to be dominant at present. 
If the YSO alignment and the elongation of the head part were originally made under the influence of HD206183 (O9.5V),
its impact might have been dominant {\em before}  the main O6 exciting star became dominant. 
This suggests that the main O6 exciting star was born quite recently compared to HD206183 (O9.5V). 
The directions almost agree with each other with the small differences of $\lesssim$20$\degr$, 
and strongly indicates that these YSOs have been formed due to the strong UV impact
of the main exciting star (O6)  and/or HD206183 (O9.5V). 

\subsection{Timescale of star formation in BRC37} 

We can estimate the speed of propagation of the sequential star formation. 
The projected distance from E1 to the IRAS source is $\sim$0.2 pc.
Assuming a period of star formation of $\sim$10$^{6}$yr, 
we derive a speed of propagation of $\sim$0.2 km s$^{-1}$.
This speed is comparable to that of the ionization front that would be created by HD206183 (O9.5V), 
if we assume a cloud density of $n$(H$_{2}$) $\sim$10$^{3}$ cm s$^{-3}$. 
A shock would precede the ionization front with a somewhat larger speed to
induce  sequential star formation, if small cores preexisted at the cloud tip
or were decoupled from the cloud tip due to the locally enhanced self-gravity
introduced by its original density fluctuation \citep[e.g.,][]{kessel03}.
On the other hand, the earlier type star HD206267 (O6) would create a significantly faster ionization front 
that could be too much to explain the speed of propagation of the sequential star formation. 
This again indicates that the YSO alignment is due to ionizing flux from HD206183 (O9.5V) 
and that HD206267 (O6)  turned on quite recently. 

The status of R1--R3 is less clear, but if we assume that R1--R3, 
located toward the $^{13}$CO clumps in the tail of the BRC37 cloud, are really YSOs, 
this would imply that star formation might occur spontaneously in the tail of the BRC37 cloud without an external agitation.
Since R1--R3 are Class I candidates,  star formation in the tail might  have begun quite recently compared to the tip region,
suggesting that star formation lags in the tail. 
Since R1--R3 are faint, their masses may be much smaller than those of the tip sources.
Such differences between the two regions could correspond to the difference 
between spontaneous and induced star formation, which lends support to
the idea of induced star formation in the tip of the BRC37 cloud. 
Alternatively, if R1--R3 are background sources, then the lack of star formation in the tail would again support this idea.

\subsection{A young brown dwarf candidate at the tip of BRC37}

Since the obvious detections of brown dwarfs \citep{rebolo95,nakajima95,oppen95},
many brown dwarfs have been  found. 
However, it is still unclear how brown dwarfs form, e.g., 
whether dwarfs form in the same manner as stars or whether the formation 
mechanism differs from those of stars and planets.
Two competing scenarios,  ejection and turbulence scenarios, are well known 
as the formation mechanisms of brown dwarfs \citep[e.g.,][]{mohsnty06}.
Recently, \cite{whitworth07} reviewed five mechanisms in detail, i.e., 
1) turbulence fragmentation of molecular clouds, 
producing very low-mass prestellar cores by shock; 
2) collapse and fragmentation of more massive prestellar cores; 
3) circumstellar disk fragmentation; 
4) premature  ejection of protostellar embryos from their natal cores; and 
5) photo-erosion of pre-existing cores overrun by HII  regions. 
They mentioned that these mechanisms are not mutually exclusive and 
that their relative importance probably depends on environment.
For the fifth mechanism, \cite{whitworth07} suggested that cores immersed 
in HII regions may be photo-eroded by the resulting ionization front and 
end up spawning brown dwarfs, from the calculations of \cite{whitworth04}.
Although \cite{whitworth04} noted that their work was concerned with cores well within 
an HII region in the vicinity of the exciting star, the lower erosion rate may be expected 
even in cores of the periphery of the HII region.

As mentioned in Section 3.2, E1 is most likely to be a young brown dwarf with a circumstellar disk, 
which is located at the outer most position among the YSO candidates.
Sequential star formation from the outside toward the inside is strongly suggested in this globule. 
Thus, E1 is very likely to have formed prior to the other candidates of T Tauri stars.

If so, it may be impossible for the brown dwarf to form  with the mechanisms 3), 
because E1 is not a companion.
Also, it may be impossible to form with the mechanism 4),
because it could be very rare that the prestellar core of E1 was ejected just to the direction 
along the globule axis (direction of sequential star formation).
If the prestellar core of E1 is formed by the mechanism 1) or 2)   \citep[e.g.,][]{kessel03}, 
it should be inevitable that this core is 
exposed to the UV radiation and is photo-eroded. 
Therefore, we conclude that it is likely that mechanism 5) follows the mechanisms 1) and 2).  
If the photo-erosion rate was not high enough to form the brown dwarf from the prestellar core 
that forms a low-mass star, the initial mass of the prestellar core could be very small.  
And if the prestellar core was very small so that it was thermally supported, 
it could be squeezed by ionization shock front.
Thus, the brown dwarf could be formed by the UV impact at the tip of BRC37.

However, it is difficult to definitely eliminate the possibility that it is 
a background  H$\alpha$ source that is severely reddened. Spectroscopy of 
the outermost source is essential to confirm whether it is really
a young brown dwarf.

\section{Summary}

We conducted near-IR ($J, H, K$$_{\rm s}$, and H$_{2}$) and optical 
($g'$, $i'$, wide H$\alpha$, $V$, $I$$_{\rm C}$, and slitless grism spectroscopy) observations of BRC37.
Our work is summarized as follows:

\begin{enumerate}
\item Ten H$\alpha$ emission stars (E1--E10) were detected around BRC37.
Six of them (E1--E6) are located at the tip and aligned along the straight line that is parallel to the axis 
of the head part of BRC37.
E2, E3, E5 and E6 are most likely to be CTTSs associated with the tip of BRC37.
\item NIR excess sources (I1--I6) were selected from the $J-H$ vs. $H-K$ color-color diagram.
One (I1) of them seems to be a Class I source embedded in the tip of BRC37, while the other sources
except I6 seem to be galaxies.  
It is possible that I6 is a foreground brown dwarf or late M-dwarf. 
\item Three red NIR sources (R1--R3), which are detectable both at $H$ and $K$$_{\rm s}$, but not at $J$,
are selected toward the $^{13}$CO clumps as possible Class I candidates. 
\item We detected H$_{2}$ emission that was probably excited by the outflows from the IRAS source,
suggesting that the IRAS source is a protostar-like object having a considerably young age. 

\item The ages of E2, E3, E5 and E6 are estimated to be $\sim$ 1.0 Myr from the evolutionary model of pre-main sequence stars. 
This suggests that they were formed at the tip in advance of the formation of the IRAS source.
The UV incident angles from the main exciting star (O6) of IC1396 and the closer exciting star (HD206183, O9.5) of BRC37, 
the axis angles of  the head part and its tip of the BRC37 cloud, and the alignment angle of YSOs are nearly identical,
indicating that the YSOs at the tip were formed under the influence of the UV radiation from the main exciting star  and/or
HD206183.
Thus, it is likely that sequential star formation has been taking place from the side of the exciting star(s) toward the 
IRAS source due to the UV impact of the exciting star(s).
\item E1 is located at the closest position to the exciting star(s) and
seems to be a young brown  dwarf that was formed by the UV impact in advance of the formation of other YSOs at the tip.
\end{enumerate}

\acknowledgements
We thank the staff at the UH 2.2-m telescope for supporting our observations.
Use of the UH 2.2-m telescope for the WFGS2 observations is supported by NAOJ.
This work was supported in part by Grant-in-Aid for Scientific Research (17039011, 19540242) 
from the Ministry of Education, Culture, Sports, Science and Technology.

\section*{Appendix}
\appendix
\section{Correction for Table 5 of OSP}
As was mentioned in Section 3.1.1,  H$\alpha$ equivalent widths (EWs)  of the H$\alpha$ emission stars 
presented in Table 5 of \cite{ogura02} were {\em not} scaled from pixels to angstroms, 
and the data presented there for BRC13 are a repetition of that for BRC12.  
Here we show  a fully corrected version of Table 5 from OSP  as an online table (Table \ref{table4}).
We recalculated EWs from the same slitless spectroscopic images as those in OSP 
with the methodology mentioned in the main text, and properly scaled.  
The coordinates of the H$\alpha$ emission stars  were also redetermined with the USNO-B1.0 catalog.
We attached remarks on spectroscopic status, e.g., contamination from near star(s), bright rim(s) 
or nebulosity, and continuum strength (weak, very weak, or invisible), 
which strongly relate to  the reliability of EW.
The stub version of Table \ref{table4} in the main text shows the data for BRC13.

In the course of this correction, we found some H$\alpha$ emission stars that have not been identified in OSP. 
We also listed these newly found stars by adding "N" to their  object numbers.

\begin{deluxetable}{lccclc}
\tabletypesize{\scriptsize}
\tablecaption{H$\alpha$ Emission Stars}
\tablewidth{0pt}
\tablehead{
\colhead{Object} & \colhead{$\alpha$} & \colhead{$\delta$} & \colhead{EW} & \colhead{Remarks} & \colhead{EW (OSP 2002)\tablenotemark{a}} \\
& \colhead{(J2000.0)} & \colhead{(J2000.0)} & \colhead{[\AA]} & & \colhead{[\AA]}
}
\startdata
E1 & 21$^{h}$40$^{m}$25$^{s}$.55\phn & +56\arcdeg36$^{'}$38$^{''}$.7\phn & invisible cont. & OSP 37-1 & invisible cont. \\
E2 & 21$^{h}$40$^{m}$25$^{s}$.97\phn & +56\arcdeg36$^{'}$32$^{''}$.1\phn & 14.9 & OSP 37-2 & 18.4\\
E3 & 21$^{h}$40$^{m}$26$^{s}$.77\phn & +56\arcdeg36$^{'}$23$^{''}$.5\phn & 36.4 & OSP 37-3 & 40.9\\
E4 & 21$^{h}$40$^{m}$27$^{s}$.43\phn & +56\arcdeg36$^{'}$20$^{''}$.0\phn & \nodata\tablenotemark{b} &   & not visible \\
E5 & 21$^{h}$40$^{m}$28$^{s}$.69\phn & +56\arcdeg36$^{'}$09$^{''}$.4\phn & 87.4\tablenotemark{c}  & OSP 37-7 & 78.8\tablenotemark{c} \\
E6 & 21$^{h}$40$^{m}$28$^{s}$.03\phn & +56\arcdeg36$^{'}$05$^{''}$.9\phn & invisible cont. & OSP 37-6, [Sll]$\lambda$6731 & invisible cont. \\
& & & &  [Nll]$\lambda$6584, [Ol]$\lambda$6300\\
E7 & 21$^{h}$40$^{m}$13$^{s}$.67\phn & +56\arcdeg40$^{'}$49$^{''}$.9\phn & 42.8 & &\nodata\\
E8 & 21$^{h}$40$^{m}$32$^{s}$.33\phn & +56\arcdeg38$^{'}$40$^{''}$.4\phn & 13.9 & OSP 37-8 & 14.4 \\
E9 & 21$^{h}$40$^{m}$42$^{s}$.27\phn & +56\arcdeg31$^{'}$13$^{''}$.2\phn & 22.5 &  & \nodata\\
E10 & 21$^{h}$40$^{m}$43$^{s}$.52\phn & +56\arcdeg31$^{'}$42$^{''}$.6\phn & invisible cont. & & \nodata\\
\enddata
\label{table1}
\tablenotetext{a}{Calculated from the spectroscopic data of \cite{ogura02} in the same manner as this work.}\\
\tablenotetext{b}{Not estimated due to contamination from a neighboring star (OSP 37-5).}
\tablenotetext{c}{Small contamination from a neighboring star and bright rim.}
\end{deluxetable}

\begin{deluxetable}{ccccccccccc}
\tabletypesize{\tiny}
\rotate
\tablecaption{Summary of YSO Identification}
\tablewidth{0pt}
\tablehead{
\colhead{Candidates} & \colhead{$g'$} & \colhead{$i'$} & \colhead{$V$} &
\colhead{$I_{\rm C}$} & \colhead{$J$} & \colhead{$H$} & \colhead{$K$} & \colhead{$A_{V}$\tablenotemark{a}} & 
\colhead{Candidate Type} & \colhead{Location}\\
& \colhead{[mag]} &\colhead{[mag]} & \colhead{[mag]} & \colhead{[mag]}
& \colhead{[mag]} & \colhead{[mag]} & \colhead{[mag]} & \colhead{[mag]} &
}
\startdata
E1 & \nodata\phn & 21.09$\pm$0.14\phn & $\ga$22.83\phn & 20.06$\pm$0.08\phn & 17.43$\pm$0.02\phn & 16.64$\pm$0.03\phn & 15.95$\pm$0.02\phn & $\la$2.00\tablenotemark{b}\phn & young brown dwarf\phn & BRC37 tip\\\\
E2 & 17.40$\pm$0.04\phn & 15.04$\pm$0.06\phn & 16.77$\pm$0.02\phn & 14.39$\pm$0.03\phn & 12.68$\pm$0.02\phn & 11.76$\pm$0.02\phn & 11.30$\pm$0.02\phn & 2.21\phn & CTTS\phn & BRC37 tip\\\\
E3 & 18.87$\pm$0.05\phn & 16.09$\pm$0.07\phn & 18.50$\pm$0.03\phn & 15.56$\pm$0.03\phn & 13.34$\pm$0.02\phn & 12.25$\pm$0.02\phn & 11.58$\pm$0.02\phn & 3.02\phn & CTTS\phn & BRC37 tip\\\\
E5 & 18.57$\pm$0.05\phn & 16.11$\pm$0.07\phn & 17.93$\pm$0.02\phn & 15.33$\pm$0.03\phn & 13.59$\pm$0.02\phn & 12.66$\pm$0.02\phn & 12.17$\pm$0.02\phn & 2.25\phn & CTTS\phn & BRC37 tip\\\\
E6 & 22.03$\pm$0.15\phn & 18.94$\pm$0.05\phn & $\ga$22.06\phn & 18.02$\pm$0.06\phn & 15.09$\pm$0.02\phn & 13.58$\pm$0.02\phn & 12.45$\pm$0.02\phn & 5.36\phn & CTTS\phn & BRC37 tip\\\\
I1 & \nodata\phn & \nodata\phn & \nodata\phn & \nodata\phn & 20.99$\pm$0.03\phn & 19.07$\pm$0.01\phn & 17.24$\pm$0.01\phn & \nodata\phn & Class I & BRC37 tip\\\\
\tableline\\
R1 & \nodata\phn & \nodata\phn & \nodata\phn & \nodata\phn & \nodata\phn & 18.85$\pm$0.11\tablenotemark{d}\phn & 16.43$\pm$0.03\tablenotemark{d}\phn & \nodata\phn & Class I ?\phn & BRC37 tail ?\\\\
R2 & \nodata\phn & \nodata\phn & \nodata\phn & \nodata\phn & \nodata\phn & 18.45$\pm$0.09\tablenotemark{d}\phn & 17.31$\pm$0.06\tablenotemark{d}\phn & \nodata\phn & Class I ?\phn & BRC37 tail ?\\\\
R3 & \nodata\phn & \nodata\phn & \nodata\phn & \nodata\phn & \nodata\phn & 18.95$\pm$0.11\tablenotemark{d}\phn & 17.71$\pm$0.07\tablenotemark{d}\phn & \nodata\phn & Class I ?\phn & BRC37 tail ?\\\\
\tableline\\
E4 & 21.17$\pm$0.13\phn & 17.30$\pm$0.09\phn & \nodata\phn & \nodata\phn & 15.28$\pm$0.02\phn & 14.02$\pm$0.02\phn & 13.34$\pm$0.02\phn & 5.10\phn & variable star\phn & background ?\phn\\\\
E7 & 20.53$\pm$0.06\phn & 18.47$\pm$0.05\phn & 20.02$\pm$0.03\phn & 17.95$\pm$0.03\phn & \nodata\phn & \nodata\phn & \nodata\phn & \nodata\phn & \nodata\phn & background\phn\\\\
E8 & 16.39$\pm$0.02\phn & 15.45$\pm$0.04\phn & 16.14$\pm$0.02\phn & 14.95$\pm$0.02\phn & 13.43$\pm$0.03\tablenotemark{c}\phn & 13.17$\pm$0.04\tablenotemark{c}\phn & 12.98$\pm$0.05\tablenotemark{c}\phn & \nodata\phn & HAEBE/classical Be ?\phn & background\phn\\
& & & & & & & & & \phn\\
E9 & 21.14$\pm$0.10\phn & 17.72$\pm$0.06\phn & 18.93$\pm$0.03\phn & 15.86$\pm$0.06\phn & \nodata\phn & \nodata\phn & \nodata\phn & \nodata\phn & variable star\phn & IC1396/background ?\phn\\
& & & & & & & & & &\phn\\
E10 & \nodata\phn & 20.28$\pm$0.10\phn & 22.10$\pm$0.11\phn & 18.59$\pm$0.04\phn & \nodata\phn & \nodata\phn & \nodata\phn & \nodata\phn & \nodata\phn & background\phn\\\\
I2 & \nodata\phn & 20.50$\pm$0.11\phn & $\ga$22.23 & 19.78$\pm$0.12\phn & 16.27$\pm$0.05\phn & 14.70$\pm$0.05\phn & 13.70$\pm$0.04\phn & \nodata\phn & galaxy\phn & background\phn\\\\
I3 & \nodata\phn & 19.91$\pm$0.08\phn & 21.90$\pm$0.11\phn & 19.22$\pm$0.04\phn & 16.79$\pm$0.02\phn & 15.89$\pm$0.02\phn & 15.29$\pm$0.02\phn & \nodata\phn & galaxy\phn & background\phn\\\\
I4 & \nodata\phn & $\ga$21.49\phn & $\ga$22.61\phn & $\ga$20.76\phn & 17.72$\pm$0.05\phn & 16.60$\pm$0.05\phn & 15.76$\pm$0.05 & \nodata\phn & galaxy ?\phn & background\phn\\\\
I5 & \nodata\phn & $\ga$22.66\phn & $\ga$22.23\phn & $\ga$21.94\phn & 18.49$\pm$0.06\phn & 17.38$\pm$0.05\phn & 16.45$\pm$0.06\phn & \nodata\phn & galaxy\phn & background\phn\\\\
I6 & \nodata\phn & 19.64$\pm$0.07 & $\ga$22.69\phn & 18.62$\pm$0.03 & 15.95$\pm$0.02\phn & 15.37$\pm$0.02\phn & 14.85$\pm$0.02 & \nodata\phn & brown dwarf/late M-dwarf ?\phn & foreground\phn\\\\
\enddata
\label{table2}
\tablenotetext{a}{Estimated from $J-H$ vs. $H-K$ diagram (see $\S$3.2).}
\tablenotetext{b}{Estimated from $J$ vs. $J-H$ diagram (see $\S$3.2).}
\tablenotetext{c}{From the 2MASS Point Source Catalog.}
\tablenotetext{d}{SIRIUS system.}
\end{deluxetable}

\begin{deluxetable}{lcccc}
\tabletypesize{\scriptsize}
\tablecaption{Directions of YSO Alignment, UV incidence, and Cloud Axes}
\tablewidth{0pt}
\tablehead{
\colhead{Direction} & \colhead{P.A.} & \colhead{Size of Elongation} & \colhead{Projected Distance} & \colhead{Ionizing Photon Flux}\\
& & & \colhead{to BRC37} & \colhead{(s$^{-1}$ cm$^{-2}$)}\phn
}
\startdata
YSO alignment\phn & $\sim$147$\degr$\phn & $\sim$1$'$\phn & \nodata\phn & \nodata\phn\\\\
UV incidence & & &\\
\hspace{0.5cm}from HD206267 (O6)\phn & 166.6$\degr$\phn & \nodata\phn & $\sim$12pc\phn & $\sim$3.7 $\times$ 10$^{8}$\phn\\
\hspace{0.5cm}from HD206183 (O9.5V)\phn & 142.7$\degr$\phn & \nodata\phn & $\sim$6pc\phn & $\sim$6.5 $\times$ 10$^{7}$\phn\\
\\
Cloud axis & & &\\
\hspace{0.5cm}Tip of head part\phn ($^{13}$CO gas)\tablenotemark{a} & $\sim$165$\degr$\phn & $\sim$0.7$'$\phn & \nodata\phn & \nodata\phn\\
\hspace{0.5cm}Head part (optical)\phn & $\sim$147$\degr$\phn & $\sim$1.5$'$\phn & \nodata\phn & \nodata\phn\\
\enddata
\label{table3}
\tablenotetext{a}{See Fig. 2 of \cite{sugitani97}.}
\end{deluxetable}

\begin{deluxetable}{lccrl}
\tablecaption{Updated and Corrected Table 5 of Ogura, Sugitani, \& Pickles (2002)}
\tablewidth{0pt}
\tablehead{
\colhead{Object} & \colhead{$\alpha$} & \colhead{$\delta$} & \colhead{EW} & \colhead{Remarks}  \\
\colhead{Number} & \colhead{(J2000.0)} & \colhead{(J2000.0)} & \colhead{(\AA)} &
}
\startdata
BRC 13 \\
\hline \\
1   & 03:00:43.77 & 60:40:04.5 &  20.4 &            \\
2   & 03:00:44.82 & 60:40:09.2 &  17.2 & double star          \\
3   & 03:00:45.33 & 60:40:39.5 &  14.5 &            \\
4   & 03:00:46.50 & 60:39:52.8 &  64.1 & comtam. fr. nr. stars        \\
5   & 03:00:50.91 & 60:40:59.6 & \nodata & contam. fr. nr. star        \\
6a  & 03:00:51.28 & 60:39:36.5 & 112.8 & double star          \\
6b  & 03:00:51.04 & 60:39:36.1 & \nodata & double star, weak cont.        \\
7   & 03:00:51.67 & 60:39:49.0 &  23.4 & contam. from brigh rim        \\
8   & 03:00:52.21 & 60:40:34.2 &  58.6 & weak cont., contam. from brigh rim      \\
9   & 03:00:52.65 & 60:40:42.5 & \nodata & contam. fr. nr. star        \\
10  & 03:00:52.71 & 60:39:31.9 &  24.1 & contam. fr. bright rims        \\
11  & 03:00:53.39 & 60:40:26.6 & \nodata & comtam. fr. nr. stars        \\
12  & 03:00:55.47 & 60:39:42.9 &  75.7 & comtam. fr. nr. stars        \\
13  & 03:00:56.03 & 60:40:26.5 &   8.2 & weak cont., comtam. fr. nr. stars      \\
14  & 03:01:02.18 & 60:39:34.5 &  72.7 & weak cont.          \\
15  & 03:01:07.40 & 60:40:40.0 &  44.8 & weak cont., comtam. from No. 16 star     \\
16  & 03:01:07.59 & 60:40:41.5 & 159.4 & weak cont., comtam. from No. 15 star     \\
17  & 03:01:08.10 & 60:39:01.9 &  36.6 & very weak cont.         \\
18  & 03:01:11.15 & 60:38:55.9 &  27.1 & very weak cont.         \\
19  & 03:01:11.51 & 60:40:56.8 &  61.5 & very weak cont.         \\
20  & 03:01:12.15 & 60:38:42.3 & \nodata & invisible cont.          \\
21  & 03:01:22.76 & 60:39:40.5 &  16.0 &            \\
22N & 03:00:54.48 & 60:39:39.2 &  11.7 &            
\enddata
\label{table4}
\end{deluxetable}

\clearpage
\begin{center}
\huge{Online Material}
\end{center}
%\documentclass{aastex}
%\begin{document}
\begin{deluxetable}{lccrl}
\tablenum{4}
\tablecaption{Updated and Corrected Table 5 of Ogura, Sugitani, \& Pickles (2002)}
\tablewidth{0pt}
\tablehead{
\colhead{Object} & \colhead{$\alpha$} & \colhead{$\delta$} & \colhead{EW} & \colhead{Remarks}  \\
\colhead{Number} & \colhead{(J2000.0)} & \colhead{(J2000.0)} & \colhead{(\AA)} &
}
\startdata
BRC 1 \\
\hline \\
1   & 23:59:42.50 & 67:22:27.8 & \nodata & invisible cont., contam. fr. nr. star      \\
2   & 23:59:43.72 & 67:25:50.0 &  26.6 & weak cont.          \\
3   & 23:59:43.99 & 67:22:46.8 & \nodata & invisible cont.          \\
4   & 23:59:47.61 & 67:23:11.2 & \nodata & H$\alpha$ ?, invisible cont.        \\
5   & 23:59:47.98 & 67:23:06.9 & \nodata & H$\alpha$ ?, weak cont.        \\
6   & 23:59:52.75 & 67:25:37.5 &  32.6 & weak cont.          \\
\hline \\
BRC 2 \\
\hline \\
1   & 00:03:51.03 & 68:33:15.8 & \nodata & H$\alpha$ ?          \\
2   & 00:03:52.32 & 68:31:58.8 & \nodata & invisible cont.          \\
3   & 00:03:54.52 & 68:33:44.6 & \nodata & contam. fr. nr. strs        \\
4   & 00:03:54.98 & 68:32:42.8 & 145.7 & very weak cont.         \\
5   & 00:03:57.12 & 68:33:46.7 &  15.6 &            \\
6   & 00:03:57.35 & 68:33:23.0 &  99.6 &            \\
7   & 00:03:58.33 & 68:34:06.5 &  13.1 &            \\
8   & 00:03:59.12 & 68:32:47.3 &  18.2 &            \\
9   & 00:04:01.70 & 68:34:13.8 &   2.8 &            \\
10  & 00:04:01.81 & 68:34:00.0 &  14.0 &            \\
11  & 00:04:01.80 & 68:34:37.4 &   5.5 &            \\
12  & 00:04:01.88 & 68:34:34.5 &  21.2 &            \\
13  & 00:04:02.32 & 68:31:36.2 & \nodata & H$\alpha$ ?          \\
14  & 00:04:02.65 & 68:34:26.6 &  24.9 &            \\
15  & 00:04:04.69 & 68:33:49.3 &  25.8 & weak cont.          \\
16  & 00:04:04.59 & 68:34:52.2 &  23.0 &            \\
17  & 00:04:05.28 & 68:33:56.0 & 136.6 & very weak cont.         \\
18  & 00:04:05.26 & 68:33:53.1 &  50.4 &            \\
19  & 00:04:05.66 & 68:33:44.3 &  94.5 &            \\
20  & 00:04:07.33 & 68:33:42.3 & \nodata & invisible cont.          \\
21  & 00:04:07.60 & 68:33:25.1 &  12.8 &            \\
22  & 00:04:11.66 & 68:33:25.4 &  46.7 &            \\
23  & 00:04:13.97 & 68:32:21.8 &  85.7 & weak cont.          \\
24  & 00:04:14.71 & 68:32:49.1 &  42.1 &            \\
25  & 00:04:15.18 & 68:33:02.0 &  16.1 &            \\
26  & 00:04:15.41 & 68:34:05.5 &  26.6 & very weak cont.         \\
27  & 00:04:21.66 & 68:30:59.6 & \nodata & H$\alpha$ ?          \\
28  & 00:04:25.36 & 68:32:31.0 &  12.3 & weak cont.          \\
29  & 00:04:29.99 & 68:31:19.9 & \nodata & H$\alpha$ ?          \\
30N & 00:03:59.88 & 68:33:41.7 &   1.4 &            \\
\hline \\
BRC 5 \\
\hline \\
1   & 02:28:44.20 & 61:31:15.4 &   6.0 &            \\
2   & 02:28:48.30 & 61:32:31.0 & \nodata & invisible cont.          \\
3   & 02:28:53.81 & 61:33:32.3 &  39.2 & weak cont.          \\
4   & 02:28:59.11 & 61:32:48.8 &  85.5 &            \\
5   & 02:29:00.18 & 61:31:57.5 & \nodata & weak cont.          \\
6   & 02:29:01.51 & 61:33:29.2 &  61.6 &            \\
7   & 02:29:05.68 & 61:32:59.2 & \nodata & contam. fr. nr. star        \\
8   & 02:29:05.31 & 61:33:22.4 & \nodata & invisible cont.          \\
9   & 02:29:08.13 & 61:34:09.9 & 198.0 & very weak cont.         \\
10  & 02:29:08.30 & 61:33:26.2 &  58.2 & very weak cont.         \\
11  & 02:29:14.16 & 61:33:25.8 &  86.5 &            \\
12N & 02:28:46.50 & 61:35:59.8 & \nodata & contam. fr. No. 13 star       \\
13N & 02:28:45.52 & 61:36:01.0 &  39.1 & contam. fr. No. 12 star       \\
\hline \\
BRC 7 \\
\hline \\
1   & 02:34:28.32 & 61:47:43.5 & \nodata & contam. fr. nr. star        \\
2   & 02:34:36.26 & 61:45:43.1 & \nodata & M-star ?          \\
3   & 02:34:41.37 & 61:46:34.0 & \nodata & contam. fr. nr. star        \\
4   & 02:34:58.17 & 61:46:48.6 & \nodata & very weak cont.         \\
5   & 02:35:00.36 & 61:46:19.1 &   1.9 &            \\
6   & 02:35:03.50 & 61:49:01.9 &   9.9 &            \\
7   & 02:35:15.30 & 61:48:03.3 &  85.2 &            \\
8   & 02:35:15.31 & 61:48:09.5 &  16.5 &            \\
9   & 02:35:16.01 & 61:47:02.4 & \nodata & invisible cont.          \\
10N & 02:34:42.47 & 61:49:41.7 & 105.0 & weak cont.          \\
11N & 02:34:41.67 & 61:50:00.7 &  29.1 &            \\
12N & 02:34:34.35 & 61:47:47.0 & \nodata & invisible cont.          \\
13N & 02:34:29.28 & 61:47:47.2 & \nodata & invisible cont.          \\
14N & 02:34:32.81 & 61:46:36.9 &  53.0 & very weak cont., contam. fr. nr. stars     \\
\hline \\
BRC 11 \\
\hline \\
1   & 02:51:32.85 & 60:03:54.0 &   6.7 &            \\
\hline \\
BRC 11NE \\
\hline \\
1   & 02:51:37.39 & 60:06:26.8 &  26.1 &            \\
2   & 02:51:47.33 & 60:07:47.2 &   6.3 &            \\
3   & 02:51:54.21 & 60:07:46.2 & \nodata & invisible cont.          \\
4   & 02:51:54.52 & 60:08:26.8 &  76.5 & contam. fr. nr. star        \\
5   & 02:51:58.70 & 60:08:06.0 &   6.6 &            \\
6   & 02:51:59.77 & 60:06:39.4 &  12.1 & weak cont.          \\
7   & 02:52:11.12 & 60:07:15.3 &  15.7 &            \\
8   & 02:52:15.03 & 60:05:18.8 & \nodata & contam. fr. nr. stars        \\
9N  & 02:51:55.02 & 60:07:17.4 & 109.0 &            \\
10N & 02:51:52.16 & 60:07:10.2 &  44.6 &            \\
11N & 02:51:51.64 & 60:07:30.6 & \nodata & contam. fr. nr. stars        \\
12N & 02:51:51.70 & 60:06:43.7 &   5.2 &            \\
\hline \\
BRC 11E \\
\hline \\
1   & 02:52:13.64 & 60:03:26.2 & 245.1 & very weak cont.         \\
2   & 02:52:15.01 & 60:05:19.4 & \nodata & contam. fr. nr. stars        \\
3N  & 02:52:14.25 & 60:03:11.7 &  55.4 &            \\
\hline \\
BRC 12 \\
\hline \\
1   & 02:54:45.06 & 60:35:38.3 &   5.1 &            \\
2   & 02:54:45.45 & 60:34:19.4 &  46.5 &            \\
3   & 02:54:46.21 & 60:34:40.9 & \nodata & invisible cont.          \\
4   & 02:54:46.29 & 60:36:14.8 & \nodata & invisible cont., contam. fr. nr. star      \\
5   & 02:54:49.51 & 60:38:26.1 &   4.5 & weak H$\alpha$ ?         \\
6   & 02:54:50.15 & 60:37:30.6 & \nodata & invisible cont.          \\
7   & 02:54:50.51 & 60:37:28.7 & \nodata & H$\alpha$ ?, weak cont.        \\
8   & 02:54:54.53 & 60:35:00.1 & \nodata & invisible cont.          \\
9   & 02:54:56.52 & 60:35:45.8 &  18.6 & contam. fr. nr. star        \\
10  & 02:54:56.85 & 60:38:14.1 & \nodata & very weak cont.         \\
11  & 02:54:58.05 & 60:35:51.0 &  57.0 & contam. fr. bright rim        \\
12  & 02:54:59.97 & 60:35:09.6 & \nodata & weak H$\alpha$ ?, contam. fr. bright rim     \\
13  & 02:55:04.51 & 60:37:27.4 & \nodata & invisible cont.          \\
14  & 02:55:07.80 & 60:36:19.9 &  33.2 & contam. fr. nr. star        \\
15  & 02:55:07.92 & 60:36:17.9 & \nodata & invisible cont.          \\
16  & 02:55:09.67 & 60:36:21.0 & \nodata & contam. fr. nr. star        \\
17  & 02:55:14.36 & 60:36:50.0 & \nodata & contam. fr. nr. star        \\
18  & 02:55:16.04 & 60:35:15.5 & \nodata & very weak cont.         \\
19  & 02:55:16.98 & 60:38:02.3 &  57.2 & contam. fr. nr. star        \\
20  & 02:55:18.61 & 60:35:21.9 &  51.7 &            \\
21  & 02:55:18.75 & 60:36:09.4 &  86.2 & contam. fr. nr. stars        \\
22  & 02:55:19.10 & 60:36:09.5 & \nodata & contam. fr. nr. stars        \\
23  & 02:55:20.23 & 60:35:14.3 & \nodata & invisible cont.          \\
24  & 02:55:20.74 & 60:38:17.7 & 100.9 & weak cont.          \\
25N & 02:54:58.72 & 60:35:48.4 &  75.5 & contam. fr. bright rim        \\
\hline \\
BRC 13 \\
\hline \\			
1   & 03:00:43.77 & 60:40:04.5 &  20.4 &            \\
2   & 03:00:44.82 & 60:40:09.2 &  17.2 & double star          \\
3   & 03:00:45.33 & 60:40:39.5 &  14.5 &            \\
4   & 03:00:46.50 & 60:39:52.8 &  64.1 & comtam. fr. nr. stars        \\
5   & 03:00:50.91 & 60:40:59.6 & \nodata & contam. fr. nr. star        \\
6a  & 03:00:51.28 & 60:39:36.5 & 112.8 & double star          \\
6b  & 03:00:51.04 & 60:39:36.1 & \nodata & double star, weak cont.        \\
7   & 03:00:51.67 & 60:39:49.0 &  23.4 & contam. from brigh rim        \\
8   & 03:00:52.21 & 60:40:34.2 &  58.6 & weak cont., contam. from brigh rim      \\
9   & 03:00:52.65 & 60:40:42.5 & \nodata & contam. fr. nr. star        \\
10  & 03:00:52.71 & 60:39:31.9 &  24.1 & contam. fr. bright rims        \\
11  & 03:00:53.39 & 60:40:26.6 & \nodata & comtam. fr. nr. stars        \\
12  & 03:00:55.47 & 60:39:42.9 &  75.7 & comtam. fr. nr. stars        \\
13  & 03:00:56.03 & 60:40:26.5 &   8.2 & weak cont., comtam. fr. nr. stars      \\
14  & 03:01:02.18 & 60:39:34.5 &  72.7 & weak cont.          \\
15  & 03:01:07.40 & 60:40:40.0 &  44.8 & weak cont., comtam. from No. 16 star     \\
16  & 03:01:07.59 & 60:40:41.5 & 159.4 & weak cont., comtam. from No. 15 star     \\
17  & 03:01:08.10 & 60:39:01.9 &  36.6 & very weak cont.         \\
18  & 03:01:11.15 & 60:38:55.9 &  27.1 & very weak cont.         \\
19  & 03:01:11.51 & 60:40:56.8 &  61.5 & very weak cont.         \\
20  & 03:01:12.15 & 60:38:42.3 & \nodata & invisible cont.          \\
21  & 03:01:22.76 & 60:39:40.5 &  16.0 &            \\
22N & 03:00:54.48 & 60:39:39.2 &  11.7 &            \\
\hline \\
BRC 14 \\
\hline \\
1   & 03:01:04.06 & 60:31:26.1 &  44.2 & weak cont.          \\
2   & 03:01:05.84 & 60:28:03.1 &  47.5 & very weak cont.         \\
3   & 03:01:06.17 & 60:30:17.9 &  83.0 & weak cont.          \\
4   & 03:01:06.56 & 60:30:36.3 & \nodata & invisible cont.          \\
5   & 03:01:07.74 & 60:29:21.9 &  38.0 &            \\
6   & 03:01:11.49 & 60:30:46.5 & 108.6 & very weak cont.         \\
7   & 03:01:13.00 & 60:28:51.7 & 102.1 & contam. fr. nr. star        \\
8   & 03:01:13.39 & 60:29:32.1 &  19.8 & weak cont.          \\
9   & 03:01:13.42 & 60:29:44.2 & \nodata & invisible cont.          \\
10  & 03:01:16.12 & 60:29:47.1 & \nodata & contam. fr. nr. star        \\
11  & 03:01:16.34 & 60:28:40.0 &  87.0 & very weak cont.         \\
12  & 03:01:17.06 & 60:29:23.2 &  13.5 &            \\
13  & 03:01:17.16 & 60:30:46.2 & \nodata & invisible cont.          \\
14  & 03:01:17.73 & 60:29:32.0 & \nodata & contam. fr. nr. star        \\
15  & 03:01:18.58 & 60:29:54.6 & \nodata & invisible cont.          \\
16  & 03:01:18.76 & 60:29:56.4 & \nodata & H$\alpha$ ?, contam. fr. nr. star      \\
17  & 03:01:19.91 & 60:30:11.6 &  38.3 & weak cont., contam. fr. nr. star      \\
18  & 03:01:20.26 & 60:30:02.2 & \nodata & contam. fr. nr. stars        \\
19  & 03:01:20.67 & 60:29:54.6 & \nodata & contam. fr. nr. star        \\
20  & 03:01:20.61 & 60:29:31.8 & \nodata & contam. fr. nr. stars        \\
21  & 03:01:20.58 & 60:29:28.0 & \nodata & invisible cont.          \\
22  & 03:01:20.86 & 60:29:46.5 & \nodata & contam. fr. nr. stars        \\
23  & 03:01:21.16 & 60:29:44.3 & 107.0 & contam. fr. nr. star        \\
24  & 03:01:21.22 & 60:30:10.4 & \nodata & contam. fr. nr. star        \\
25  & 03:01:21.59 & 60:28:56.6 &  22.1 &            \\
26  & 03:01:22.38 & 60:30:02.9 & \nodata & contam. fr. nr. star and bright rim     \\
27  & 03:01:23.48 & 60:30:50.4 & \nodata & invisible cont.          \\
28  & 03:01:23.82 & 60:29:58.5 & \nodata & contam. fr. bright rim        \\
29  & 03:01:24.01 & 60:30:42.5 & \nodata & contam. fr. nr. star        \\
30  & 03:01:24.73 & 60:30:09.6 & \nodata & weak cont., contam. fr. bright rim      \\
31  & 03:01:25.57 & 60:29:39.0 &  14.0 &            \\
32  & 03:01:26.39 & 60:30:53.9 &  10.1 &            \\
33  & 03:01:27.20 & 60:30:56.8 &  29.8 & weak cont.          \\
34  & 03:01:27.41 & 60:30:39.4 & \nodata & contam. fr. nr. star and bright rim     \\
35  & 03:01:29.29 & 60:31:13.6 &  55.9 &            \\
36  & 03:01:30.43 & 60:31:02.5 &  99.1 & very weak cont.         \\
37  & 03:01:31.00 & 60:30:26.9 & \nodata & invisible cont., contam. fr. nr. stars      \\
38  & 03:01:31.19 & 60:30:20.8 & \nodata & weak H$\alpha$ ?, contam. fr. nr. star     \\
39  & 03:01:33.92 & 60:27:45.9 &   9.8 &            \\
40  & 03:01:34.33 & 60:30:08.5 &  14.5 &            \\
41  & 03:01:36.36 & 60:29:06.1 &  30.9 & contam. fr. nr. star        \\
42  & 03:01:36.91 & 60:31:00.1 &  22.6 & contam. fr. No. 44 star       \\
43  & 03:01:37.03 & 60:29:41.1 &   5.3 & M-star ?          \\
44  & 03:01:36.96 & 60:31:01.2 & \nodata & contam. fr. No. 42 star       \\
45  & 03:01:37.54 & 60:28:54.0 & \nodata & invisible cont.          \\
46  & 03:01:43.27 & 60:28:51.3 & \nodata & H$\alpha$ ?          \\
47  & 03:01:49.88 & 60:28:50.7 &  31.3 & weak cont.          \\
48N & 03:01:20.31 & 60:29:49.2 &   6.2 &            \\
49N & 03:01:23.51 & 60:31:51.0 & \nodata & invisible cont., near upper edge       \\
50N & 03:01:05.22 & 60:31:55.5 &  59.2 & cor. fr. DSS II R \\
\hline \\
BRC 27 \\
\hline \\
1   & 07:03:52.49 & -11:26:16.9 &  17.5 &            \\
2   & 07:03:52.70 & -11:23:13.5 & \nodata & double star          \\
3   & 07:03:53.22 & -11:24:03.8 & \nodata & invisible cont.          \\
4   & 07:03:53.72 & -11:24:28.7 &  30.2 &            \\
5   & 07:03:54.98 & -11:25:14.6 &  13.3 &            \\
6   & 07:03:56.37 & -11:25:41.4 & \nodata & H$\alpha$ ?, double star, contam. fr. nr. star    \\
7   & 07:03:57.11 & -11:24:32.9 & \nodata & H$\alpha$ ?          \\
8   & 07:04:02.86 & -11:23:37.4 &  15.6 & contam. fr. No. 9 star       \\
9   & 07:04:02.93 & -11:23:39.1 &   8.1 & contam. fr. No. 8 star       \\
10  & 07:04:03.07 & -11:23:50.8 &  36.4 & contam. fr. nr. star and bright rim     \\
11  & 07:04:04.07 & -11:26:35.4 &  23.2 & weak cont.          \\
12  & 07:04:04.24 & -11:23:55.9 & 166.5 & weak cont.          \\
13  & 07:04:04.57 & -11:25:55.0 &   3.8 & H$\alpha$ ?          \\
14  & 07:04:04.68 & -11:23:40.0 &  10.4 &            \\
15  & 07:04:05.19 & -11:23:13.5 &  33.2 & contam. fr. nr. stars        \\
16  & 07:04:05.91 & -11:23:58.9 &  25.0 & bad pix. at Ha        \\
17  & 07:04:06.02 & -11:23:16.0 & \nodata & contam. fr. No. 19 star       \\
18  & 07:04:06.43 & -11:23:36.3 & 109.0 & weak cont., contam. fr. nr. star      \\
19  & 07:04:06.55 & -11:23:16.6 &  14.0 & contam. fr. nr. stars (incl. No. 17 star) and bright rim \\
20  & 07:04:06.68 & -11:26:08.5 &  54.1 &            \\
21  & 07:04:07.96 & -11:23:11.8 & \nodata & H$\alpha$ ?, contam. fr. nr. stars      \\
22  & 07:04:08.02 & -11:23:54.9 &  38.9 &            \\
23  & 07:04:08.16 & -11:23:09.8 & 231.2 & weak cont.          \\
24  & 07:04:09.24 & -11:24:38.3 &  44.0 & weak cont.          \\
25  & 07:04:09.94 & -11:23:16.6 &  53.4 &            \\
26  & 07:04:10.93 & -11:23:27.6 & \nodata & H$\alpha$ ?, contam. fr. nr. star      \\
27  & 07:04:11.95 & -11:24:22.9 &  62.1 & weak cont.          \\
28  & 07:04:12.75 & -11:24:03.5 & 182.2 & weak cont., contam. fr. nr. star      \\
29  & 07:04:13.52 & -11:24:55.9 &  16.9 &            \\
30  & 07:04:13.91 & -11:24:41.8 &  12.9 &            \\
31  & 07:04:14.25 & -11:23:17.1 &  30.2 & near image edge         \\
32  & 07:04:14.27 & -11:23:37.3 & \nodata & weak cont., near image edge       \\
33N & 07:04:02.23 & -11:25:43.0 &   8.8 &            \\
\hline \\
BRC 30 \\
\hline \\
1   & 18:18:35.35 & -13:45:48.0 &  15.2 &            \\
2   & 18:18:35.51 & -13:48:25.9 &  92.2 & weak cont.          \\
3   & 18:18:36.33 & -13:49:06.3 & \nodata & very weak cont.         \\
4   & 18:18:36.52 & -13:49:30.1 &   6.6 & M-star ?          \\
5   & 18:18:37.26 & -13:48:26.2 &  24.7 & M-star ?, weak cont.        \\
6   & 18:18:37.27 & -13:50:13.7 &  65.5 & weak cont.          \\
7   & 18:18:37.30 & -13:45:57.0 & \nodata & contam. fr. nr. star        \\
8   & 18:18:37.65 & -13:46:06.4 & \nodata & contam. fr. nr. star        \\
9   & 18:18:37.69 & -13:46:08.4 & \nodata & contam. fr. nr. star        \\
10  & 18:18:38.25 & -13:47:40.3 & \nodata & invisible cont.          \\
11  & 18:18:38.24 & -13:51:17.1 &  62.3 &            \\
12  & 18:18:38.35 & -13:47:33.4 & \nodata & contam. fr. nr. star        \\
13  & 18:18:38.60 & -13:49:24.3 & \nodata & invisible cont.          \\
14  & 18:18:38.72 & -13:47:10.3 & \nodata & contam. fr. nr. stars        \\
15  & 18:18:38.81 & -13:46:44.3 & \nodata & saturated           \\
16  & 18:18:39.33 & -13:46:54.4 & \nodata & contam. fr. nr. stars        \\
17  & 18:18:39.41 & -13:48:08.2 & \nodata & invisible cont.          \\
18  & 18:18:39.74 & -13:47:48.1 & \nodata & weak cont., contam. fr. nr. stars      \\
19  & 18:18:40.08 & -13:47:00.8 &   5.0 &            \\
20  & 18:18:40.11 & -13:46:31.3 &  13.3 & contam. fr. nr. star        \\
21  & 18:18:41.25 & -13:49:14.2 & \nodata & very weak cont., contam. fr. nr. stars     \\
22  & 18:18:41.34 & -13:45:44.8 &   8.6 &            \\
23  & 18:18:41.40 & -13:48:35.8 & \nodata & invisible cont.          \\
24  & 18:18:41.51 & -13:48:10.9 & 136.3 & very weak cont.         \\
25  & 18:18:41.48 & -13:45:53.7 & \nodata & H$\alpha$ ?          \\
26  & 18:18:41.63 & -13:50:13.2 &  89.9 &            \\
27  & 18:18:41.66 & -13:47:12.6 & \nodata & contam. fr. nr. star        \\
28  & 18:18:41.87 & -13:47:21.4 & \nodata & contam. fr. nr. stars (incl. No. 31 star)    \\
29  & 18:18:41.93 & -13:48:59.8 &   3.4 &            \\
30  & 18:18:42.02 & -13:49:42.3 &  85.1 &            \\
31  & 18:18:42.17 & -13:47:22.4 & \nodata & contam. fr. nr. stars (inc. No. 28 star)    \\
32  & 18:18:42.33 & -13:49:54.2 &  27.4 & weak cont.          \\
33  & 18:18:42.41 & -13:46:10.9 & \nodata & weak cont., contam. fr. nebula       \\
34  & 18:18:42.53 & -13:47:15.5 & \nodata & contam. fr. nr. stars (incl. No. 36)     \\
35  & 18:18:42.82 & -13:47:27.9 &  22.2 & contam. fr. nr. star        \\
36  & 18:18:43.02 & -13:47:15.2 & \nodata & [NII], contam. fr. nr. stars (incl. No. 34)    \\
37  & 18:18:43.02 & -13:49:50.6 &  27.2 &            \\
38  & 18:18:43.19 & -13:46:39.2 &  19.4 &            \\
39  & 18:18:43.28 & -13:50:03.4 &  63.3 &            \\
40  & 18:18:43.80 & -13:46:09.5 &  40.0 & contam. fr. nr. star        \\
41  & 18:18:43.82 & -13:48:04.3 & \nodata & invisible cont., contam. fr. nr. star      \\
42  & 18:18:44.01 & -13:46:38.0 & \nodata & weak cont., contam. fr. nr. stars      \\
43  & 18:18:44.20 & -13:48:17.6 &   7.1 &            \\
44  & 18:18:44.44 & -13:50:58.5 & \nodata & weak cont., contam. fr. nebulosity       \\
45  & 18:18:44.63 & -13:48:00.5 & \nodata & weak cont., contam. fr. nr. star      \\
46  & 18:18:44.84 & -13:47:48.5 & \nodata & contam. fr. nr. star        \\
47  & 18:18:44.88 & -13:49:03.7 & \nodata & double star, contam. fr. nebulosity       \\
48  & 18:18:45.04 & -13:47:52.2 & \nodata & contam. fr. nr. star        \\
49  & 18:18:45.18 & -13:48:37.4 &  14.1 &            \\
50  & 18:18:45.30 & -13:48:40.1 & \nodata & H$\alpha$ ?          \\
51  & 18:18:45.65 & -13:46:56.9 &   7.8 & contam. fr. nr. star        \\
52 & 18:18:45.86 & -13:46:53.8 &   6.3 & contam. fr. nr. star        \\
53  & 18:18:46.02 & -13:47:30.2 &  43.1 & contam. fr. nebulosity         \\
54  & 18:18:46.05 & -13:47:10.9 & \nodata & H$\alpha$ ?, contam. fr. nr. star      \\
55  & 18:18:46.29 & -13:48:09.1 & \nodata & invisible cont.          \\
56  & 18:18:46.31 & -13:49:26.7 & \nodata & very weak cont., contam. fr. nr. star     \\
57  & 18:18:46.80 & -13:49:32.0 &   8.5 &            \\
58  & 18:18:46.86 & -13:49:16.5 &   7.8 &            \\
59  & 18:18:47.02 & -13:49:29.2 &  16.4 &            \\
60  & 18:18:47.96 & -13:48:36.3 & 112.4 & contam. fr. nr. stars        \\
61  & 18:18:48.10 & -13:48:20.2 &  11.4 & weak cont., contam. fr. nebulosity       \\
62  & 18:18:48.20 & -13:49:08.3 &  31.5 &            \\
63  & 18:18:48.39 & -13:45:49.8 & \nodata & H$\alpha$ ?          \\
64  & 18:18:48.64 & -13:47:49.7 &  17.3 & contam. fr. nebulosity         \\
65  & 18:18:48.91 & -13:48:32.3 &  20.2 & weak cont., contam. fr. nebulosity       \\
66  & 18:18:49.76 & -13:47:46.5 & \nodata & H$\alpha$ ?          \\
67  & 18:18:49.88 & -13:46:59.7 &   7.8 &            \\
68  & 18:18:52.10 & -13:48:10.3 &  14.0 & contam. fr. nr. star        \\
69  & 18:18:52.24 & -13:46:39.1 &  20.2 &            \\
70  & 18:18:52.39 & -13:46:58.5 &   6.6 &            \\
71  & 18:18:52.81 & -13:48:36.9 &  85.0 & weak cont.          \\
72  & 18:18:53.05 & -13:46:18.6 &  39.7 &            \\
73 & 18:18:54.52 & -13:47:17.3 &   8.7 &            \\
74  & 18:18:55.04 & -13:46:44.6 &  25.4 &            \\
75  & 18:18:55.30 & -13:46:51.8 & \nodata & contam. fr. nr. star        \\
76  & 18:18:57.75 & -13:44:42.4 & \nodata & contam. fr. nr. star        \\
77  & 18:18:58.52 & -13:42:45.9 &  21.1 & contam. fr. nr. star        \\
78  & 18:18:58.50 & -13:48:28.6 &   8.0 &            \\
79  & 18:18:59.18 & -13:45:58.5 & \nodata & invisible cont.          \\
80  & 18:19:01.08 & -13:45:20.9 &  38.9 & very weak cont.         \\
81  & 18:19:01.46 & -13:44:42.6 & \nodata & invisible cont.          \\
82  & 18:19:01.60 & -13:45:16.3 &   4.5 &            \\
\hline \\
BRC 31 \\
\hline \\
1   & 20:50:36.94 & 44:21:41.0 &  49.9 &            \\
2   & 20:50:36.98 & 44:20:50.1 &  29.3 &            \\
3   & 20:50:37.37 & 44:20:53.3 &   7.5 &            \\
4   & 20:50:37.47 & 44:21:16.2 &  60.0 &            \\
5   & 20:50:40.45 & 44:20:22.1 &  66.3 &            \\
6   & 20:50:40.54 & 44:20:51.0 &  27.3 & contam. fr. bright rim        \\
7   & 20:50:42.72 & 44:21:53.4 &  39.1 & weak cont., contam. fr. No. 8 star     \\
8   & 20:50:42.81 & 44:21:55.5 & \nodata & double star, both show H$\alpha$ emission      \\
9   & 20:50:44.97 & 44:20:59.4 &  28.9 & weak cont.          \\
10  & 20:50:45.40 & 44:21:49.9 &  32.7 & very weak cont., contam. fr. bright rim     \\
11  & 20:50:48.59 & 44:21:05.0 &  14.4 &            \\
12  & 20:50:48.63 & 44:20:25.4 & \nodata & H$\alpha$ ?          \\
13  & 20:50:48.70 & 44:20:53.7 &  27.2 &            \\
14  & 20:50:50.40 & 44:21:39.0 & \nodata & H$\alpha$ ?          \\
15  & 20:50:50.96 & 44:23:46.8 &  75.4 & weak cont., contam. fr. bright rim      \\
16  & 20:50:51.26 & 44:21:32.0 &  31.4 & weak cont.          \\
17  & 20:50:51.72 & 44:20:46.4 & \nodata & invisible cont.          \\
18  & 20:50:53.17 & 44:23:53.8 &  14.1 &            \\
19  & 20:50:53.60 & 44:22:12.7 & \nodata & star ?, H$\alpha$ ?        \\
20  & 20:50:53.53 & 44:24:27.2 &  12.5 &            \\
21  & 20:50:53.58 & 44:21:01.4 &   9.0 &            \\
22  & 20:50:53.62 & 44:24:30.9 &  22.1 &            \\
23  & 20:50:53.85 & 44:21:40.3 & \nodata & contam. fr. nr. stars        \\
24  & 20:50:53.73 & 44:21:18.9 & \nodata & contam. fr. No. 25 star       \\
25  & 20:50:53.94 & 44:21:18.9 & \nodata & contam. fr. No. 24 star       \\
26  & 20:50:54.89 & 44:22:58.5 &  91.8 &            \\
27  & 20:50:54.86 & 44:20:13.1 &  31.5 &            \\
28  & 20:50:55.58 & 44:24:24.9 & 121.6 & very weak cont., contam. fr. No. 20 star    \\
29  & 20:50:58.74 & 44:23:17.7 &  99.0 &            \\
30  & 20:51:02.14 & 44:20:09.7 & 101.5 &            \\
31  & 20:51:03.10 & 44:24:03.3 & \nodata & H$\alpha$ ? M-star ?        \\
32  & 20:51:03.68 & 44:24:16.1 &  60.4 &            \\
33N & 20:50:59.00 & 44:19:54.9 &  35.6 & contam. fr. nr. star        \\
34N & 20:50:59.47 & 44:19:39.7 &  72.4 &            \\
35N & 20:50:48.80 & 44:19:23.9 &  16.7 &            \\
36N & 20:50:46.09 & 44:19:10.4 &   8.6 &            \\
\hline \\
BRC 33 \\
\hline \\
1   & 21:34:19.65 & 57:30:02.6 &  53.4 &            \\
2   & 21:34:20.68 & 57:30:49.0 &   3.3 &            \\
3   & 21:34:49.11 & 57:31:25.1 &  66.2 & very weak cont.         \\
\hline \\
BRC 34 \\
\hline \\
1   & 21:33:29.26 & 58:02:50.9 &  43.0 & weak cont.          \\
2   & 21:33:55.58 & 58:01:18.7 & \nodata & H$\alpha$ ?          \\
\hline \\
BRC 37 \\
\hline \\
1   & 21:40:25.59 & 56:36:38.9 & \nodata & invisible cont.          \\
2   & 21:40:26.01 & 56:36:32.4 &  18.4 &            \\
3   & 21:40:26.80 & 56:36:23.8 &  40.9 &        \\
4   & 21:40:27.03 & 56:36:31.4 & \nodata & H$\alpha$  ?       \\
5   & 21:40:27.37 & 56:36:21.9 & \nodata & M-star           \\
6   & 21:40:28.08 & 56:36:06.1 & \nodata & invisible cont.          \\
7   & 21:40:28.73 & 56:36:09.9 &  78.8 & contam. fr. nr. star and bright rim        \\
8   & 21:40:32.35 & 56:38:40.6 &  14.4 &            \\
\hline \\
BRC 38 \\
\hline \\
1   & 21:40:25.61 & 58:14:28.4 &  22.2 & contam. fr. nr. star        \\
2   & 21:40:27.31 & 58:14:21.3 &  59.3 & contam. fr. nr. star        \\
3   & 21:40:28.00 & 58:15:14.4 &  20.7 &            \\
4   & 21:40:31.55 & 58:17:55.8 & \nodata & H$\alpha$ ?          \\
5   & 21:40:36.59 & 58:13:46.1 &   4.0 &            \\
6   & 21:40:36.94 & 58:14:38.1 &  63.3 &            \\
7   & 21:40:37.07 & 58:15:03.1 &  29.8 &            \\
8   & 21:40:40.35 & 58:13:42.9 & \nodata & H$\alpha$ ?          \\
9   & 21:40:41.20 & 58:15:11.4 &  26.1 &            \\
10  & 21:40:41.59 & 58:14:25.8 &  14.8 &            \\
11  & 21:40:44.88 & 58:15:03.5 &  75.7 & weak cont.          \\
12  & 21:40:48.05 & 58:15:37.8 &  19.0 &            \\
13  & 21:40:48.79 & 58:15:00.3 & \nodata & H$\alpha$ ?          \\
14  & 21:40:48.92 & 58:15:11.1 & \nodata & invisible cont., contam. fr. nr. star      \\
15  & 21:40:49.11 & 58:17:09.4 &  22.2 & weak cont.          \\
16  & 21:41:01.81 & 58:15:25.4 & \nodata & H$\alpha$ ?          \\
17N & 21:40:38.86 & 58:13:03.1 & 100.2 &            \\
\hline \\
BRC 39 \\
\hline \\
1   & 21:45:50.30 & 57:26:49.7 & \nodata & H$\alpha$ ?          \\
2   & 21:46:01.61 & 57:29:38.3 &   3.1 &            \\
3   & 21:46:07.12 & 57:26:31.6 &  13.0 & [SII], [NII], [OI] ?        \\
4   & 21:46:25.96 & 57:28:28.9 & \nodata & contam. fr. nr. star        \\
5N  & 21:45:54.08 & 57:28:18.5 &   9.3 & contam. fr. nr. star        \\
\hline \\
BRC 43 \\
\hline \\
1   & 22:47:35.90 & 58:04:33.3 &   7.0 & contam. fr. nr. stars        \\
2   & 22:47:43.47 & 58:05:26.1 &  26.9 &            \\
3   & 22:47:44.57 & 58:03:54.3 &  32.7 & bad pix., contam. fr. nr. sta      \\
4   & 22:47:46.40 & 58:04:01.1 & \nodata & weak cont., contam. fr. nr. star      \\
5   & 22:47:48.22 & 58:04:25.6 &  28.2 & very weak cont.         \\
6   & 22:47:49.29 & 58:04:51.1 &  48.6 & very weak cont.         \\
7   & 22:47:55.07 & 58:03:22.8 & \nodata & H$\alpha$ ?, contam. fr. nr. star      \\
8   & 22:47:56.41 & 58:03:17.7 &  62.6 & very weak cont., contam. fr. nr. star     \\
9   & 22:48:00.24 & 58:02:41.8 & \nodata & contam. fr. nr. star        \\
10  & 22:48:00.27 & 58:02:30.4 & \nodata & double star, H$\alpha$/rim emission ?       \\
11  & 22:48:03.73 & 58:05:12.3 &  47.7 & very weak cont.         \\
12  & 22:48:06.55 & 58:04:37.9 &  14.8 &            \\
13  & 22:48:10.46 & 58:04:11.4 &  10.8 & bad pix.          \\
14  & 22:48:11.53 & 58:03:25.0 & \nodata & H$\alpha$ ?          \\
\hline \\
BRC 44 \\
\hline \\
1   & 22:28:19.01 & 64:13:54.0 &  55.8 &            \\
2   & 22:28:21.00 & 64:14:13.2 &   4.2 & reflection nebula ?         \\
3   & 22:28:41.77 & 64:13:11.6 & \nodata & contam. fr. nr. star        \\
4   & 22:28:43.54 & 64:13:19.1 & \nodata & double star          \\
5   & 22:28:43.98 & 64:13:26.0 &  50.6 &            \\
6   & 22:28:44.12 & 64:13:31.2 &  20.1 & very weak cont.         \\
7   & 22:28:44.74 & 64:13:12.0 & \nodata & contam. fr. nr. stars        \\
8   & 22:28:45.31 & 64:13:05.6 &  22.5 & bad pix.          \\
9   & 22:28:45.48 & 64:13:23.0 & \nodata & contam. fr. No. 11 star       \\
10  & 22:28:45.77 & 64:13:24.9 & \nodata & contam. fr. Nos. 5 and 12 stars     \\
11  & 22:28:45.96 & 64:13:22.1 &  53.5 & contam. fr. No. 9 star       \\
12  & 22:28:46.24 & 64:13:25.6 & \nodata & contam. fr. No. 5 star       \\
13  & 22:28:47.12 & 64:13:14.3 &   9.5 &            \\
\hline \\
BRC 82 \\
\hline \\
1   & 16:47:08.23 & -41:16:08.9 &  29.7 &            \\
2   & 16:47:08.79 & -41:16:15.2 &   7.9 &            \\
3   & 16:47:13.33 & -41:16:39.7 & \nodata & contam. fr. nr. stars        \\
4   & 16:47:14.08 & -41:16:00.4 & \nodata & H$\alpha$ ?          \\
5   & 16:47:16.51 & -41:15:26.6 & \nodata & contam. fr. nr. star        \\
6   & 16:47:16.50 & -41:16:29.9 &  18.5 & contam. fr. nr. star        \\
7   & 16:47:18.66 & -41:16:31.2 & \nodata & contam. fr. nr. stars        \\
8   & 16:47:18.81 & -41:16:10.3 & \nodata & H$\alpha$ ?          \\
9   & 16:47:19.68 & -41:16:14.0 &  15.5 & contam. fr. nr. stars        \\
10  & 16:47:21.70 & -41:15:30.6 &  52.2 & very weak cont., contam. fr. nr. star     \\
11  & 16:47:22.94 & -41:13:54.3 &  52.4 & contam. fr. nr. star and bright rim     \\
12  & 16:47:23.04 & -41:15:54.5 &  73.6 & weak cont.          \\
13  & 16:47:24.08 & -41:16:07.1 &   3.2 & contam. fr. nr. star        \\
14  & 16:47:24.17 & -41:16:21.6 &   6.3 & contam. fr. nr. star and bright rim     \\
15  & 16:47:27.68 & -41:16:29.6 & 158.4 & contam. fr. nr. stars        \\
16  & 16:47:33.28 & -41:13:58.3 &   2.7 &            \\
\hline \\
BRC 89 \\
\hline \\
1   & 18:09:48.07 & -24:06:05.6 & \nodata & H$\alpha$ ?          \\
2   & 18:09:50.03 & -24:04:36.2 & \nodata & very weak cont., contam. fr. bright rim     \\
3   & 18:09:50.46 & -24:05:20.3 & \nodata & H$\alpha$ ?          \\
4   & 18:09:51.25 & -24:05:51.9 &  75.8 &            \\
5   & 18:09:53.92 & -24:07:31.3 &  18.7 & contam. fr. No. 6 star       \\
6   & 18:09:53.93 & -24:07:32.4 & \nodata & contam. fr. No. 5 star       \\
7   & 18:09:54.22 & -24:06:06.1 &   3.6 &            \\
8   & 18:09:54.62 & -24:06:15.9 & 219.4 & very weak cont.         \\
9   & 18:09:54.72 & -24:05:43.8 &  46.0 &            \\
10  & 18:09:55.60 & -24:06:58.0 & \nodata & contam. fr. No. 12 star       \\
11  & 18:09:55.85 & -24:06:09.7 &  71.6 & contam. fr. nr. star        \\
12  & 18:09:56.11 & -24:06:58.5 & \nodata & contam. fr. No. 10 star       \\
13  & 18:09:56.28 & -24:06:48.9 &  25.2 & contam. fr. nr. star        \\
14  & 18:09:56.53 & -24:06:29.8 & \nodata & contam. fr. nr. star        \\
15  & 18:09:57.49 & -24:06:07.1 &  30.1 & very weak cont.         \\
16  & 18:09:57.81 & -24:07:16.5 &  91.5 & weak cont.          \\
17  & 18:09:57.94 & -24:06:57.1 &  13.8 & weak cont.          \\
18  & 18:09:58.87 & -24:06:52.7 & \nodata & contam. fr. nr. star        \\
19  & 18:10:00.16 & -24:07:00.3 &  23.9 & very weak cont.         \\
20  & 18:10:00.22 & -24:05:19.4 &  21.4 &            \\
21  & 18:10:00.29 & -24:06:28.8 &   7.2 &            \\
22  & 18:10:00.27 & -24:06:51.6 &  59.7 & very weak cont.         \\
23  & 18:10:03.48 & -24:07:41.6 &  27.7 & weak cont.          \\
\hline \\
S140  \\
\hline \\
1   & 22:18:47.71 & 63:18:17.9 & \nodata & contam. fr. nr. stars        \\
2   & 22:18:48.52 & 63:16:40.4 &  13.6 &            \\
3   & 22:18:49.55 & 63:18:56.0 & \nodata & H$\alpha$ ?          \\
4   & 22:18:58.89 & 63:18:11.8 &  94.8 &            \\
5   & 22:18:59.87 & 63:18:52.5 & \nodata & H$\alpha$ ?          \\
6   & 22:19:03.44 & 63:18:00.3 & \nodata & H$\alpha$ ?., contam. fr. bright rim      \\
7   & 22:19:09.78 & 63:17:21.1 &  26.4 &            \\
8   & 22:19:16.82 & 63:17:22.0 & 163.5 & very weak cont.         \\
\hline \\
Cep B \\
\hline \\
1   & 22:56:37.97 & 62:39:51.1 &  69.7 &            \\
2   & 22:56:38.12 & 62:40:58.7 & 104.7 & very weak cont.         \\
3   & 22:56:39.33 & 62:38:15.5 &  80.1 &            \\
4   & 22:56:39.58 & 62:38:43.1 &  17.2 & very weak cont.         \\
5   & 22:56:39.93 & 62:41:37.1 &  14.6 & M-star ?          \\
6   & 22:56:43.54 & 62:38:07.5 & \nodata & invisible cont.          \\
7   & 22:56:45.33 & 62:41:15.8 &   8.0 & M-star ?          \\
8   & 22:56:47.79 & 62:38:14.0 &  22.4 &            \\
9   & 22:56:48.02 & 62:38:40.2 & 125.6 & very weak cont.         \\
10  & 22:56:48.23 & 62:39:11.1 &  62.3 & very weak cont.         \\
11  & 22:56:49.54 & 62:41:10.0 &  21.4 &            \\
12  & 22:56:49.77 & 62:40:30.1 &  81.3 & very weak cont.         \\
13  & 22:56:51.41 & 62:38:55.8 & \nodata & invisible cont.          \\
14  & 22:56:52.40 & 62:40:59.6 &  63.6 & contam. fr. nr. star        \\
15  & 22:56:53.87 & 62:41:17.7 &  19.0 &            \\
16  & 22:56:54.65 & 62:38:57.8 &  16.8 & weak cont., contam. fr. bright rim      \\
17  & 22:56:56.11 & 62:39:30.8 & \nodata & invisible cont.          \\
18  & 22:56:57.83 & 62:40:14.0 &  59.8 &            \\
19  & 22:56:58.65 & 62:40:56.0 & \nodata & H$\alpha$ ?          \\
20  & 22:56:59.68 & 62:39:20.2 &  76.4 &            \\
21  & 22:57:00.22 & 62:39:09.4 &  24.9 & weak cont.          \\
22  & 22:57:01.88 & 62:37:52.1 & \nodata & invisible cont.          \\
23  & 22:57:02.63 & 62:41:48.7 & \nodata & contam. fr. nr. star        \\
24  & 22:57:02.93 & 62:41:14.9 &   4.9 &            \\
25  & 22:57:04.31 & 62:38:21.1 & \nodata & contam. fr. neigh. stars        \\
26  & 22:57:07.86 & 62:41:33.2 &  23.1 & weak cont.          \\
27  & 22:57:10.82 & 62:40:51.0 & \nodata & invisible cont., contam. fr. bright rim      \\
28  & 22:57:11.48 & 62:38:14.1 &  39.6 & very weak cont.         \\
29  & 22:57:12.10 & 62:41:48.1 & \nodata & contam. fr. No. 30 star       \\
30  & 22:57:13.26 & 62:41:49.3 & \nodata & contam. fr. No. 29 star       \\
31  & 22:57:14.11 & 62:41:19.8 & \nodata & double star, both show H$\alpha$ emission      \\
32  & 22:57:19.38 & 62:40:22.5 & \nodata & H$\alpha$ ?, contam. fr. bright. rim      \\
33  & 22:57:27.04 & 62:41:07.9 &   6.4 &            \\
34N & 22:57:04.93 & 62:38:23.2 &  14.2 &            \\
35N & 22:57:05.91 & 62:38:18.4 &  10.1 &            \\
36N & 22:56:36.14 & 62:36:45.9 & \nodata & invisible cont.          \\
37N & 22:56:35.29 & 62:39:07.8 &   8.1 &            \\
\hline \\
LW Cas \\
\hline \\
1   & 02:57:03.22 & 60:41:39.8 &  14.7 &            \\
2   & 02:57:08.73 & 60:42:30.2 &  33.9 & weak cont.          \\
3   & 02:57:13.80 & 60:40:59.3 &  30.8 &            \\
4   & 02:57:15.70 & 60:38:49.3 & \nodata & M-star ?          \\
5   & 02:57:20.82 & 60:41:27.4 &  55.4 & very weak cont.         \\
6   & 02:57:27.08 & 60:41:51.4 &   8.2 & contam. fr. No. 7 star       \\
7   & 02:57:28.11 & 60:41:49.2 &  36.0 & contam. fr. No. 6 star       \\
8   & 02:57:28.34 & 60:42:33.0 & \nodata & H$\alpha$ ?          \\
9   & 02:57:29.53 & 60:42:28.1 & \nodata & double star          \\
10  & 02:57:32.81 & 60:41:29.7 &  48.6 & very weak cont.         \\
11  & 02:57:37.67 & 60:39:58.6 & \nodata & invisible cont.          \\
12  & 02:57:40.53 & 60:39:07.5 &   9.9 & very weak cont.         \\
13N & 02:57:51.23 & 60:40:00.9 &  53.7 & very weak cont.         \\
\hline \\
Ori I-2N \\
\hline \\
1   & 05:37:51.12 & -1:36:17.0 & \nodata & H$\alpha$ ?, weak cont.        \\
2   & 05:37:59.11 & -1:36:56.3 &   7.0 &            \\
3   & 05:38:02.32 & -1:36:34.4 & \nodata & H$\alpha$ ?, weak cont.        \\
4   & 05:38:02.58 & -1:34:39.5 & \nodata & M-star ?          \\
5N  & 05:37:51.41 & -1:36:42.3 &   6.7 &            \\
\enddata
%\label{table4}
\end{deluxetable}
%\end{document}

\end{document}